\documentclass[%
 reprint,
superscriptaddress,
nofootinbib,
bibnotes,
 amsmath,amssymb,
 aps,
]{revtex4-1}

\usepackage{latexsym,mathrsfs,bbm}
\usepackage{tikz}
\usetikzlibrary{backgrounds,fit,decorations.pathreplacing,calc}
\usepackage{graphicx}
\usepackage{dcolumn}
\usepackage{braket}
\usepackage{gensymb}
\usepackage{bm}
\usepackage{amsmath}
\usepackage{amsthm}
\usepackage{float}
\usepackage{cancel}
\usepackage{mathtools}
\usepackage{color}
\usepackage{csquotes}
\usepackage[utf8]{inputenc}
\usepackage{thmtools}

\declaretheorem[]{theorem}

\usepackage{diagbox}
\usepackage{adjustbox}

\theoremstyle{plain}

\newtheorem{Cor}{Corollary}

\def\id{\mathbbm{1}}   

\DeclareMathOperator{\Tr}{Tr}
\newcommand{\bT}{\bm{\theta}}
\newcommand{\bTe}{\bm{\theta}^{0}}
\newcommand{\rT}{\hat{\rho}_{\bT}}
\newcommand{\rTe}{\hat{\rho}_{\bT^{0}}}
\newcommand{\pps}{p_{\bT}^{\mathrm{ps}}}

\usepackage{hyperref}

\allowdisplaybreaks

\begin{document}

\preprint{APS/123-QED}

\title{Quantum Learnability is Arbitrarily Distillable}

\author{Joe H. Jenne}
\affiliation{%
 Cavendish Laboratory, Department of Physics, University of Cambridge, Cambridge CB3 0HE, United Kingdom
}%

\author{David R. M. Arvidsson-Shukur}
\affiliation{%
Hitachi Cambridge Laboratory, J. J. Thomson Avenue, CB3 0HE, Cambridge, United Kingdom
}%
\affiliation{%
 Cavendish Laboratory, Department of Physics, University of Cambridge, Cambridge CB3 0HE, United Kingdom
}%
\affiliation{%
Research Laboratory of Electronics, Massachusetts Institute of Technology, Cambridge, Massachusetts 02139, USA
}%

\date{\today}

\begin{abstract}
Quantum learning (in metrology and machine learning) involves  estimating unknown parameters $\bm{\theta} = ( \theta_1, \dots, \theta_M )$ from measurements of  quantum states $\rT$. The quantum Fisher information matrix can bound the average amount of information learnt about $\bT$ per experimental trial. In several scenarios, it is advantageous to concentrate information in as few states as possible. Here, we present two ``go-go'' theorems proving that negativity, a narrower nonclassicality concept than noncommutation, enables unbounded and lossless distillation of Fisher information about multiple parameters in quantum learning.
\end{abstract}

\maketitle


The use of experimental data to estimate unknown parameters $\bm{\theta} = ( \theta_1, \theta_2, \dots, \theta_M )$ is a quintessential task in metrology and many machine-learning algorithms. In quantum metrology and quantum machine-learning, nonclassical phenomena are used to improve the learning of $\bm{\theta}$ based on measurements of quantum states $\rT$ \cite{Giovanetti11, Maccone13, Szczykulska16, Kiani21}. Here, we show that \textit{negativity} \cite{ArvShuk20}, a narrower concept than noncommutation, enables unbounded and lossless distillation of information about multiple parameters in quantum learning.

A common  measure of an experiment's usefulness in learning (estimating) multiple unknown parameters $\bm{\theta} $ is the Fisher information matrix $I(\bm{\theta})$ \cite{Braunstein94, Liu19, Albarelli20}. $I(\bm{\theta})$  quantifies the average information learned about  $\bm{\theta}$ from one experimental trial. The  covariance matrix of a locally unbiased estimator $\bm{\theta}^{\mathrm{e}}$ is lower-bounded by the Cramér-Rao inequality: $\Sigma(\bm{\theta}^{\mathrm{e}}) \geq \left[ N I(\bm{\theta})  \right]^{-1}$, where $N$ is the number of independent experimental trials \cite{Rao92, Cramer16}. Theoretically, the learning task is then to adjust the experimental input state and final measurement to optimize the Fisher information matrix and to minimize the estimator's risk with respect to some risk function \cite{Ballester04, Fujiwara07, Genoni13, Humphreys13, Pezze17,  Chen17}. However, such results are not necessarily representative of optimal experimental strategies---especially in quantum experiments.

Whilst a theorist aims to optimize the  Fisher information, an experimentalist must manage her \textit{cost} \cite{Liuzzo18, Lipka18}. Recent works, theoretical and practical, have focused on limiting experimental costs associated with the measurement and post-processing of  output states.  \textit{Weak-value} amplification \cite{ Dressel14, Harris17,  Xu20} and postselected metrology \cite{ArvShukur19-2, Lupu20} allows the rate of output states per unit time to be lowered whilst a significant fraction of the information about a single parameter $\theta_1$ is retained. This enables detectors to operate at lower intensities and can, if the postselection is experimentally  \textit{cheap}, reduce temporal overheads associated with measurements and postprocessing. The protocols cannot increase the information content, but can reduce the experimental costs of accessing it. A major shortcoming of most previous information-distillation protocols is that they require perfect knowledge of all-but-one experimental parameter---an often unrealistic setting.\footnote{Initial studies of  weak-value amplification with specific forms of multiparameter unitaries are given in \cite{Vella19, Xia20, Ho21}. }

Given the important role of multiparameter learning in quantum metrology and quantum machine learning, a generalization of these results is crucial for both practical and foundational reasons. A generalization will help facilitate postselected metrology in diverse experiments, where several parameters are (partially) unknown, as well as in quantum machine-learning, where the overhead associated with the postprocessing of output data can be monumental. From a foundational perspective, a generalization could provide useful knowledge about the nature of negativity and noncommutation as quantum resources,  as well as about the fundamental limits of encoding information in quantum states.

In this Article, we provide this generalization. First, we review theoretical results, establishing that scalar risk functions based on the  \textit{quantum} Fisher information matrix are suitable objects to minimize, when optimizing quantum learning. Second, we derive a formula for the distilled (postselected) quantum Fisher information matrix. Third, we use a Kirkwood-Dirac quasiprobability distribution \cite{Kirkwood33, Dirac45, Yunger18} (a diversified cousin of the Wigner function) to find classical and nonclassical bounds on the entries in the quantum Fisher information matrix (Thm. \ref{Th:NCQFIM}).\footnote{In this work, if the experiment is described by an operationally defined  quasiprobability distribution (see below) that does not equal a classical probability distribution, we call the experiment nonclassical.} We prove that the presence of negative quasiprobabilities allows the quantum Fisher information matrix to take anomalous entries, outside the classical bounds. Fourth, we design a quantum-learning protocol in which the useful information in an arbitrarily large number of states $\rT$ is distilled into an arbitrarily small number of states $\rT^{\mathrm{ps}}$ (Thm. \ref{Th:DivPSLearn}). Our protocol is lossless: no information is wasted in the distillation (postselection) procedure. Fifth, we discuss how our results can be applied to improve quantum learning in the presence of imperfect detectors  or postprocessing costs.

\section{Preliminaries}

Consider an experiment with finite and discrete outcomes $k$ with corresponding probabilities $p(k | \bT )$. The Fisher information matrix is defined as
\begin{equation}
\label{Eq:ClasFish}
I_{i,j}(\bT) = \sum_k p(k | \bT) \left\{ \partial_i \log[  p(k | \bT) ] \right\} \left\{ \partial_j \log[  p(k | \bT) ] \right\} ,
\end{equation}
where $\partial_i \equiv \frac{\partial}{\partial\theta_i}$ \cite{bCover06}. The Fisher information matrix lower-bounds the covariance matrix $\Sigma(\bm{\theta}^{\mathrm{e}})$ via the Cramér-Rao inequality: $\Sigma(\bm{\theta}^{\mathrm{e}}) \geq \left[ N I(\bm{\theta})  \right]^{-1}$.  Choosing a positive, real, $M \times M$ weight matrix $W$, introduces a scalar Cramér-Rao bound: 
\begin{equation}
\label{Eq:ScalCB}
s(\Sigma(\bm{\theta}^{\mathrm{e}}), W) \equiv  \Tr{ \left[ W \Sigma(\bm{\theta}^{\mathrm{e}})  \right]} \geq  \frac{1}{N} \Tr{ \left[ W I^{-1} (\bm{\theta}) \right]} .
\end{equation}
If, e.g., $W = \id$ and $\bm{\theta}^{\mathrm{e}}$ is an unbiased estimator, the scalar risk function $s(\Sigma(\bm{\theta}^{\mathrm{e}}), W)$  equals the sum of the individual mean-square errors of the parameters in $\bm{\theta}^{\mathrm{e}}$. See \cite{Albarelli20} for a review. For unbiased,  or ``reasonable'', estimators $\bm{\theta}^{\mathrm{e}}$ and $N \to \infty$,  Ineq. \eqref{Eq:ScalCB} is saturated \cite{Lehmann06}. In what follows, we shall assume these conditions, such that $s(\Sigma(\bm{\theta}^{\mathrm{e}}), W) \equiv s(I (\bm{\theta}), W) = \Tr{ \left[ W I^{-1} (\bm{\theta}) \right]} /N$.

From a learnability perspective, it is often useful to consider the most informative experiment that extracts (Fisher) information from quantum states $\rT $:
\begin{align}
s^{(\mathrm{MI})}(\rT, W)  \equiv s\left( \max_{\mathcal{M}}  I(\bT), W \right)
= \frac{1}{N} \min_{\mathcal{M}} \Tr{ \left[ W I^{-1} (\bm{\theta}) \right]}.
\end{align}
Here, $\mathcal{M}$ is the set of all possible measurements. 
 
The Fisher information matrix is upper-bounded by the  quantum Fisher information matrix  \cite{Helstrom67, Liu19, Albarelli20}: $I(\bT) \leq 	\mathcal{I}(\bT | \rT )$.  The quantum Fisher information matrix is defined by
\begin{equation}
	\mathcal{I}_{i,j} (\bT | \rT ) = \Tr\left( \hat{L}_j \partial_i \rT \right).
	\label{Eq:QFIM}
\end{equation} 
Here, $\hat{L}_j$ is the logarithmic derivative operator, which is not uniquely defined \cite{Liu19}.   It can be defined using a symmetric logarithmic derivative (SLD), $2 \partial_i \rT =  \hat{L}^{\mathrm{(SLD)}}_i \rT +\rT \hat{L}^{\mathrm{(SLD)}}_i  $, or with a right logarithmic derivative (RLD), $\partial_i \rT = \rT \hat{L}^{\mathrm{(RLD)}}_i$. In the multiparameter scenario ($M> 1$), noncommutation often forbids measurements such that $I_{i,j}(\bT) = \mathcal{I}_{i,j}(\bT)$ for all $i,j$. Thus, $I(\bT) \leq 	\mathcal{I}(\bT | \rT )$ cannot commonly be saturated. Either the symmetric-logarithmic-derivative or the right-logarithmic-derivative quantum Fisher information matrix can give a bound that lies closer to the achievable bound.  For pure states  $\hat{L}^{\mathrm{(SLD)}}_i = 2 \hat{L}^{\mathrm{(RLD)}}_i = 2 \partial_i \rT $, and the symmetric-logarithmic-derivative quantum Fisher information matrix [Eq. \eqref{Eq:QFIM}] is  
\begin{equation}
	\mathcal{I}_{i,j}(\bT | \psi_{\bT} ) = 4\Re \left[ \braket{\partial_i\psi_{\bT}|\partial_j\psi_{\bT}} - \braket{\partial_i\psi_{\bT}|\psi_{\bT}}\braket{\psi_{\bT}|\partial_j\psi_{\bT}} \right],
	\label{Eq:liuqfim}
\end{equation}
where  $\rT\equiv \ket{\psi_{\bT}}\bra{\psi_{\bT}}$ \cite{Liu19}. In this theoretical proof-of-principle study, we  proceed with an investigation of pure states and the symmetric-logarithmic-derivative quantum Fisher information matrix. An investigation of distilled quantum learning  in the presence of noise is left for an upcoming paper.

The quantum Fisher information matrix yields a scalar Cramér-Rao bound \cite{Albarelli20}:
\begin{align}
\label{Eq:ScalQCB}
s^{(\mathrm{MI})}(\rT, W) \geq \frac{1}{N}  \Tr{ \left[ W \mathcal{I}^{-1} (\bm{\theta}) \right]}.
\end{align}
It is this bound that (directly or indirectly) leads  quantum machine-learning algorithms to optimize the quantum Fisher information matrix of their subroutines \cite{Abbas20, Haug21, Meyer21}. However, Eq. \eqref{Eq:ScalQCB} ``only'' provides a lower bound on $s^{(\mathrm{MI})}(\rT, W)$. Consequently, it is reasonable to ask: How good a measure of learnability is the quantum Fisher information matrix? From an information theoretic perspective, the answer \cite{Albarelli19, Carollo19} is given by
\begin{equation}
\label{Eq:QuantScal}
\frac{1}{N} \Tr{ \left[ W \mathcal{I}^{-1} (\bm{\theta}) \right]} \leq  h(\bT, W) \leq (1+\mathcal{Q})\frac{1}{N} \Tr{ \left[ W \mathcal{I}^{-1} (\bm{\theta}) \right]},
\end{equation}
where $h(\bT, W)$ is  Holevo's  lower bound  of the Cramér-Rao inequality \cite{Holevo77}. The ``geometric quantumness'' measure $\mathcal{Q}$ (see Appendix \ref{App:PSQuant}) satisfies $0 \leq \mathcal{Q} \leq 1$.  Generally, it is hard to calculate 
$h(\bT, W)$ (see \cite{Albarelli20} for the exact form). Nevertheless, for pure states, $s^{(\mathrm{MI})}(\rT, W) =h(\bT, W) $ \cite{Matsumoto02}. 

For the purpose of the theoretical pure-state investigation in this work, the formulae above can be summarized as
\begin{equation}
\label{Eq:ScalQCBPure}
\frac{1}{N} \Tr{ \left[ W \mathcal{I}^{-1} (\bm{\theta}) \right]} \leq  s^{(\mathrm{MI})}(\rT, W)  \leq  2 \frac{1}{N} \Tr{ \left[ W \mathcal{I}^{-1} (\bm{\theta}) \right]}.
\end{equation}
Within a factor of $2$, $ \mathcal{I}(\bm{\theta}) $ sets $s^{(\mathrm{MI})}(\rT, W)$.
This constitutes our main motivation for focusing on $\mathcal{I} (\bm{\theta})$ as a measure of quantum learnability. Further, empirical motivation, can be found in Refs. \cite{Abbas20, Haug21, Meyer21}.

\section{Postselected Quantum Fisher Information Matrix}

\begin{figure}
\includegraphics[scale=0.25]{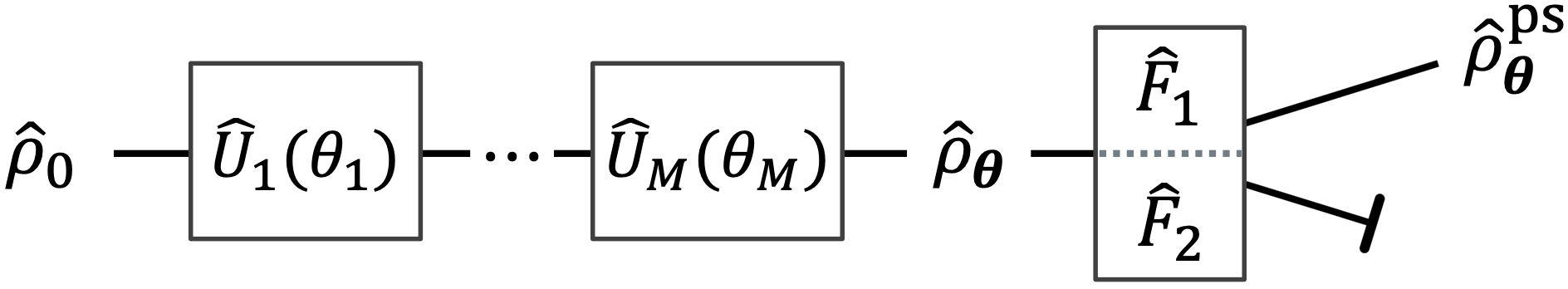}
\caption{\textbf{Preparation and distillation of quantum states.}  First, $M$ unknown parameters $\bT$ are encoded in the initial state $\hat{\rho}_0$ by the unitary $\hat{U}(\bT)$: $\hat{\rho}_0 \rightarrow \rT$. Second, the encoded state $\rT$ is past through a postselective measurement $\{\hat{F}_1 =\hat{F} , \hat{F}_2 =\hat{1} - \hat{F} \}$. The postselection is used to discard the quantum states unless outcome $\hat{F} = \hat{K}^{\dagger} \hat{K}$ happens. Third, the  experiment outputs the distilled states $\rT^{\textrm{ps}} = \hat{K} \rT \hat{K}^{\dagger} / \pps$ with success probability $\pps =\mathrm{Tr}(\hat{F}\rT) $.  }
\label{fig:EncodingCircuit}
\end{figure}

 Here, we consider  an experiment  where an initial state, $\rho_0$, is evolved sequentially  by a series of $M$ unitary operators, $\hat{U}(\bT) \equiv \prod_{m=M}^1 \hat{U}_m(\theta_m)$: $\hat{\rho}_0 \rightarrow \rT \equiv \hat{U}(\bT) \hat{\rho}_0 \hat{U}^{\dagger}(\bT)$, and then subject to a  postselective measurement $\{ \hat{F}_1 = \hat{F}, \hat{F}_2 = \hat{1} - \hat{F} \}$. $\hat{F}_{i}$ need not be projective. The experiment  is depicted in Fig. \ref{fig:EncodingCircuit}. We assume a discrete Hilbert space of dimension $D$ and that  $\hat{U}_m(\theta_m)$ satisfies Stone's theorem on one-parameter unitary groups \cite{Stone32} such that $
	\hat{U}_m(\theta_m)=e^{i\theta_m \hat{A}_m} \; \forall m \in 1,\ldots,M$.\footnote{Experiments where $\hat{U}_m(\theta_m) \neq e^{i\theta_m \hat{A}_m}$ can often be transformed into the required form via an artificial reparametrization.}
The Hermitian generators $\hat{A}_m$ are in general noncommuting.

We now present a formula for the quantum Fisher information matrix of the distilled fraction of the output states  in Fig. \ref{fig:EncodingCircuit}. These output states are given by $\ket{\psi_{\bT}^{\mathrm{ps}}} \equiv \hat{K} \ket{\psi_{\bT}} / \sqrt{\pps}$, where $\pps = \Tr[\hat{F} \rT]$ is the probability of successful postselection and $\hat{K}$ is the Kraus operator that sets the postselection: $\hat{F} =  \hat{K}^{\dagger}  \hat{K}$. In Appendix \ref{App:PSQFIM}, we evaluate Eq. \eqref{Eq:QFIM} for $\ket{\psi^{\mathrm{ps}}_{\bT}}$ and  we find that
\begin{align}
	\mathcal{I}_{i,j} (\bT | \psi^{\mathrm{ps}}_{\bT} ) = & 4\Re \bigg\{ \frac{1}{\pps} \Tr\left[ \hat{F} \hat{\tilde{A}}_j \rT \hat{\tilde{A}}_i \right] \nonumber \\
	& - \frac{1}{(\pps)^2} \Tr\left[\hat{F} \rT \hat{\tilde{A}}_i \right]\Tr\left[ \hat{F} \hat{\tilde{A}}_j \rT \right] \bigg\} .
	\label{Eq:PSqfim}
\end{align}
Here,    $
\hat{\tilde{A}}_m \equiv   \left[ \prod_{i=M}^{m+1} \hat{U}_i(\theta_i) \right] \hat{A}_m  \left[ \prod_{j=m+1}^{M} \hat{U}^{\dagger}_j(\theta_j)  \right]
$ for $m<M$. For $m=M$, $\hat{\tilde{A}}_M = \hat{A}_M$.
 The eigenspectra of $\hat{\tilde{A}}_m$ and $\hat{A}_m$ are identical.

\section{Quasiprobabilistic analysis}

We use  quasiprobabilistic techniques to bound $\mathcal{I}_{i,j} (\bT | \psi^{\mathrm{ps}}_{\bT} )$ with respective to classical and quantum statistics. Quasiprobability distributions are mathematical objects that behave similar to probability distributions: They sum to unity, and marginalizing over all but one of the arguments yields a classical probability distribution. However, individual quasiprobabilities can be nonclassical by having values outside $[0,1]$. This enables the distributions to describe noncommuting quantum mechanics. The best-known quasiprobability distribution is the complementary- and continuous-variable Wigner function \cite{Wigner32}. However, most of modern quantum information research is framed in terms of discrete systems, e.g., systems of qubits; and observables of interest are not necessarily complementary. 

 The complex-valued Kirkwood-Dirac (KD) quasiprobability distribution \cite{Kirkwood33, Dirac45}  is a relative of the Wigner function  that describes,  straightforwardly,  discrete systems---even qubits. The KD distribution has recently illuminated quantum effects in  weak-value amplification \cite{Steinberg95, Dressel15, Yunger18}, measurement disturbance \cite{ Hofmann11, Dressel12, Dressel15, Monroe20}, tomography  \cite{Johansen07, Lundeen11, Lundeen12, Bamber14, Thekkadath16}, quantum chaos \cite{Swingle16, Yunger18, Yunger18-2, Yunger19, Landsman19, Razieh19}, metrology \cite{ ArvShukur19-2, Lupu20},  thermodynamics \cite{Levy19, Lostaglio20}, and the foundations of quantum mechanics \cite{Griffiths84, Goldstein95, Hartle04, Hofmann11, Hofmann12-2, Hofmann14, Hofmann15, Hofmann16, Halliwell16, Stacey19}. By optimizing a formula with respect to a classical (real and non-negative) and a quantum (complex) Kirkwood-Dirac distribution, classical and quantum bounds can be found, respectively. Below we deploy this technique.

A KD distribution represents a quantum state $\hat{\rho}$ in terms of $k \geq 2$  sets of measurement operators. Equation \eqref{Eq:PSqfim} can be decomposed naturally in terms of a KD distribution defined by a discrete $\hat{\rho}$ and $k=3$ sets of measurement operators. Two sets are composed of the  projectors onto the subspaces of distinct eigenvalues of $\hat{\tilde{A}}_i$ and $\hat{\tilde{A}}_j$, and one set contains the postselection measurement operators:
\begin{align*}
	& \left\{ \hat{\Pi}^{(i)}_k \,  : \, \hat{\Pi}^{(i)}_k  \hat{\tilde{A}}_i = a^{(i)}_k  \hat{\Pi}^{(i)}_k  \right\} ,\\
	&	\left\{ \hat{\Pi}^{(j)}_l \,  : \, \hat{\Pi}^{(j)}_l  \hat{\tilde{A}}_j = a^{(j)}_l  \hat{\Pi}^{(j)}_l  \right\} ,\\
	 & \left\{ \hat{F}_1=\hat{F}, \hat{F}_2 = \hat{1} - \hat{F} \right\}.
\end{align*}
We  order the eigenvalues of $\hat{\tilde{A}}_i$ and $\hat{\tilde{A}}_j$ ascendingly: $a_1^{(i)} \leq\cdots\leq a_D^{(i)}$, and define the spectral eigengap $\Delta a^{(i)} \equiv a_D^{(i)} - a_1^{(i)}$ etc. We can now define our operational KD distribution with respect to the operators above:
\begin{equation}
\left\{ q^{\hat{\rho}}_{k,l,m} \right\} \equiv \left\{ \Tr \left[  \hat{\Pi}^{(i)}_k \hat{F}_m  \hat{\Pi}^{(j)}_l \hat{\rho} \right] \right\} .
\end{equation}

The KD distribution obeys an analogue of Bayes' Theorem \cite{Johansen07,  Yunger18}. Consequently, we can define a distribution that corresponds to $\{ q^{\hat{\rho}}_{k,l,m} \}$ conditioned on the postselection yielding outcome $\hat{F}$:
\begin{align}
\left\{ Q^{\hat{\rho}}_{k,l} \right\} 
& \equiv \left\{ \dfrac{ q^{\hat{\rho}}_{k,l,m=1} }{\sum\limits_{k,l,m =1 } q^{\hat{\rho}}_{k,l,m}} \right\}  = \left\{   \Tr \left[  \hat{\Pi}^{(i)}_k \hat{F}  \hat{\Pi}^{(j)}_l \hat{\rho} \right] / \pps \right\} .
\label{Eq:PSKD}
\end{align}
When $\{ Q^{\hat{\rho}}_{k,l} \}$ is classical, all $| Q^{\hat{\rho}}_{k,l} | \leq 1$. Negative quasiprobabilities allow the denominators of Eq. \eqref{Eq:PSKD} to approach $0$ even for finite numerators. Then, $| Q^{\hat{\rho}}_{k,l} |$ can be arbitrarily large. Such negativity, an example below shows, enables $\left| \mathcal{I}_{i,j} (\bT | \psi^{\mathrm{ps}}_{\bT} ) \right|$ to be anomalously large, compared to experiments described by classical distributions. This can increase distilled states' multiparameter information to nonclassically large values.
\begin{theorem}[Necessary condition for anomalous postselected quantum Fisher information matrix]
\label{Th:NCQFIM}
Suppose that a postselected quantum Fisher information matrix has some entry $\left| \mathcal{I}_{i,j} (\bT | \psi^{\mathrm{ps}}_{\bT} ) \right|> \Delta a^{(i)} \Delta a^{(j)}$. Then, an underlying KD distribution $\left\{ Q^{\hat{\rho}}_{k,l} \right\} $ necessarily contains at least one negative value.
\end{theorem}
\textit{Proof of Thm. \ref{Th:NCQFIM}:} We prove this theorem by contradiction. First, we  use Distribution \eqref{Eq:PSKD} to recast Eq. \eqref{Eq:PSqfim}:

\begin{align}
\label{Eq:KDQFIM}
	\mathcal{I}_{i,j} (\bT | \psi^{\mathrm{ps}}_{\bT} ) = & 4\Re \bigg\{ \sum\limits_{k,l} a^{(i)}_k a^{(j)}_l  Q^{\rT}_{k,l} \nonumber \\
	& - \bigg( \sum\limits_{k^{\prime},l^{\prime}} a^{(i)}_{k^{\prime}}   Q^{\rT}_{k^{\prime},l^{\prime}} \bigg) \bigg( \sum\limits_{k^{\prime \prime},l^{\prime \prime}}  a^{(j)}_{l^{\prime \prime}}  Q^{\rT}_{k^{\prime \prime},l^{\prime \prime}} \bigg)  \bigg\} .
\end{align}
Equation \eqref{Eq:KDQFIM} is a quantum extension of a covariance, where $Q^{\rT}_{k,l}$ replaces classical joint probabilities. Second, we assume that $\{ Q^{\rT}_{k,l} \}$ is classical. Third, ignoring the specific form of  $\{ Q^{\rT}_{k,l} \}$, we maximize and minimize Eq. \eqref{Eq:KDQFIM} over all classical distributions. When $Q^{\rT}_{k,l} \in [0,1]$ and $i \neq j$,  Eq. \eqref{Eq:KDQFIM} has the form of ($4$ times) a classical covariance with maximum and minimum values $ \Delta a^{(i)} \Delta a^{(j)}$ and $- \Delta a^{(i)} \Delta a^{(j)}$, respectively.\footnote{Applying the Cauchy-Schwarz inequality to a covariance of random variables $X$ and $Y$ yields $|\mathrm{Cov}(X,Y)| \leq \sqrt{\mathrm{Var}(X) \mathrm{Var}(Y)} $.} When $Q^{\rT}_{k,l} \in [0,1]$ and $i = j$, Eq. \eqref{Eq:KDQFIM}   is upper-bounded  by $(\Delta a^{(i)})^2$  and lower-bounded by $0$ \cite{ArvShukur19-2}.  Per definition, an anomalous QFIM entry breaks these  bounds, such that the assumption of a classical distribution $\{ Q^{\rT}_{k,l} \}$ cannot be satisfied. Consequently, if $ \left| \mathcal{I}_{i,j} (\bT | \psi^{\mathrm{ps}}_{\bT} ) \right| > \Delta a^{(i)} \Delta a^{(j)}  $, then $\{ Q^{\rT}_{k,l} \}$ is nonclassical. The form of Eq.  \eqref{Eq:KDQFIM} implies that any nonreal values cancel. Thus, the nonclassicality must be in the form of negativity.$\square$


An immediate corollary follows:
\begin{Cor}
\label{Cor:NonComQFIM}
	In a classically commuting theory, a theory in which operators commute, the quantum Fisher information matrix satisfies $\left| \mathcal{I}_{i,j} (\bT | \psi^{\mathrm{ps}}_{\bT} ) \right| \leq \Delta a^{(i)} \Delta a^{(j)}  $.
\end{Cor}
\textit{Proof:} Reference \cite{ArvShuk20} proves that noncommutation is necessary for nonclassical KD distributions.\footnote{In fact, noncommutation is necessary, but not sufficient, for KD nonclassicality \cite{ArvShuk20}.} The corollary thus follows from Thm. \ref{Th:NCQFIM}.$\square$

\section{Distilling quantum learnability}

If an underlying KD distribution possesses negative values, it is possible to use postselection to distil quantum Fisher information such that $\mathcal{I}_{i,j} (\bT | \psi^{\mathrm{ps}}_{\bT} ) > \Delta a^{(i)} \Delta a^{(j)}  $. However, $\mathcal{I}^{-1} (\bT | \psi^{\mathrm{ps}}_{\bT} ) $ bounds $s^{(\mathrm{MI})}(\psi^{\mathrm{ps}}_{\bT}, W) $ via matrix inequalities [Ineqs. \eqref{Eq:ScalQCBPure}], and it is generally hard to know which $\mathcal{I}_{i,j} (\bT | \psi^{\mathrm{ps}}_{\bT} )$ would be beneficial to amplify. Furthermore, setting a postselection operator $\hat{F}$ to optimize one entry in $\mathcal{I}(\bT | \psi^{\mathrm{ps}}_{\bT} )$ could have a detrimental effect on another entry.   Below, we show that it is possible to chose  $\hat{F}$ such that $\det{\mathcal{I}}(\bT | \psi^{\mathrm{ps}}_{\bT} ) \rightarrow \infty$, and $s^{(\mathrm{MI})}(\psi^{\mathrm{ps}}_{\bT}, W) \rightarrow 0$.  The price to pay for larger portions of $\mathcal{I}(\bT | \psi^{\mathrm{ps}}_{\bT} )$ is smaller success chances $\pps$. First, we provide a guiding example of two-parameter estimation of a postselected qubit.  Then, we present general theory. 

\subsection{Example}
Consider a qubit in an initial state $\ket{\psi_0} = \ket{0}$. The quantum circuit of interest is parametrized by two parameters $\bT = (\theta_1, \theta_2)$ and represented by the unitary $\hat{U}(\bT) = e^{i (\hat{\sigma}_x+\hat{\sigma}_z)  \theta_2 /\sqrt{2} } e^{i \hat{\sigma}_x \theta_1} $, where $\hat{\sigma}_k$ is the $k^{\mathrm{th}}$ Pauli operator. The quantum Fisher information matrix of the output state $\ket{\psi_{\bT}} = \hat{U}(\bT) \ket{0}$ is
\begin{equation}
\mathcal{I}(\bT | \psi_{\bT} ) = 
\begin{pmatrix}
   4 \; \; \; & 2\sqrt{2}  \\
    2\sqrt{2} \; \; \; & 3 - \cos{(4\theta_1)}
 \end{pmatrix} .
\end{equation}
We assume that our initial guess of $\bT$ is off by $1/10$ for both $\theta_1$ and $\theta_2$: $\bTe = (\theta_1 + \frac{1}{10}, \theta_2 + \frac{1}{10})$. We set the Kraus operator to $\hat{K} = (\frac{1}{\sqrt{10}} - 1) \rTe + \hat{1} $. The probability of a successful postselection is given by $\pps = \Tr[\hat{K}^{\dagger} \hat{K} \rT] \approx 1/10$. Moreover, the postselected (distilled) quantum Fisher information matrix is given by 
\begin{equation}
\label{Eq:PSMat}
\mathcal{I}(\bT | \psi_{\bT}^{\mathrm{ps}} ) \approx 10
\begin{pmatrix}
   4 \; \; \;& 2\sqrt{2}  \\
    2\sqrt{2} \; \; \; & 3 - \cos{(4\theta_1)}
 \end{pmatrix}  .
\end{equation}
All entries of $ \mathcal{I}(\bT | \psi_{\bT}^{\mathrm{ps}} ) $  break their classical maximum of $\Delta a^{(i)} \Delta a^{(j)} = 4$.
By reducing the number of quantum states that will reach the final detector by a factor of ten, we have also achieved a tenfold increase of the information content of the remaining states.

\subsection{General theory}

Here, we outline how to achieve a diverging  quantum Fisher information matrix in the general scenario. We give the following theorem
\begin{theorem}[Arbitrary distillation of quantum learnability]
\label{Th:DivPSLearn}
For a sufficiently accurate initial estimate, the theoretically attainable average information per trial about the unknown parameter vector $\bT$ has no upper limit: It is possible to distill quantum states such that $\mathcal{I} (\bT | \psi^{\mathrm{ps}}_{\bT} )   \rightarrow \infty$ and  $ s^{(\mathrm{MI})}(\psi^{\mathrm{ps}}_{\bT} , W) \rightarrow 0$ in a lossless fashion.
\end{theorem}
\textit{Proof of Thm. \ref{Th:DivPSLearn}:} Our proof is constructive. We present a specific protocol that achieves the objective; other protocols might exist.  Our results assume  that we possess an \textit{initial} estimate of $\bT$, $\bTe$, that is sufficiently close to the true value: $\bT^{0} \approx \bT$.\footnote{Also in weak-value amplification and single-parameter metrology, conducting the optimal measurement generally requires a good initial estimate of the unknown parameter of interest. Moreover, many variational quantum algorithms, e.g. for quantum computational chemistry, require  good initial estimates $\bT^{0} \approx \bT$ \cite{Tang19, Grimsley19, Yordanov20, Benjamin20, Lavrijsen20, Bittel21}. } In the limit of many trials $N \rightarrow \infty$, we can always ``sacrifice'' a vanishingly small fraction of the trials to achieve such an initial estimate. $\bT^{0}$ can also be improved iteratively, suitably using a Kalman filter \cite{Zarchan00}. Defining $\delta_m \equiv \theta_m - \theta_m^0$ such that $M \delta^2 \approx 0 $, $\rTe$ is given by
\begin{align}
\label{Eq:EsrT}
\rTe \equiv \hat{U}(\bTe) \hat{\rho}_0 \hat{U}^{\dagger}(\bTe) = \rT +  \left[ \rT \, , \, \hat{D} \right] +  \mathcal{O}\left( \delta^2 \right),
\end{align}
where $\hat{D} \equiv - i \sum_{m=1}^{M} \delta_m \hat{\tilde{A}}_m$ and, as before,  $
\hat{\tilde{A}}_m \equiv   \left[ \prod_{i=M}^{m+1} \hat{U}_i(\theta_i) \right] \hat{A}_m  \left[ \prod_{j=m+1}^{M} \hat{U}^{\dagger}_j(\theta_j)  \right]
$.

We consider the setup depicted in Fig. \ref{fig:EncodingCircuit}, with postselected quantum Fisher information given by Eqs. \eqref{Eq:PSqfim} and \eqref{Eq:KDQFIM}. We set  the Kraus operator $\hat{K}$ with respect to the initial estimate of the quantum state before postselection:\footnote{This choice of $\hat{K}$ generalizes the technique used by Lupu Gladstein \textit{et al.} in  single-parameter metrology of optical qubits \cite{Lupu20}.}
\begin{equation}
\label{Eq:OptF}
\hat{K} = (t-1) \rTe +  \hat{1}  ,
\end{equation}
where $0 \leq t \leq 1$.
Physically, this choice of $\hat{K}$ generates a postselection (distillation) procedure that transmits the expected state $\rTe$ with probability $t^2$ and transmits fully any state orthogonal to $\rTe$. Substituting $\hat{K}$ and $\rTe$ into $\mathcal{I}_{i,j} (\bT | \psi^{\mathrm{ps}}_{\bT} )$ [Eq. \eqref{Eq:PSqfim}] yields
\begin{align}
\label{Eq:DivQFIM}
\mathcal{I}_{i,j} (\bT | \psi^{\mathrm{ps}}_{\bT} ) =    \frac{ 1}{ t^2  } \mathcal{I}_{i,j} (\bT | \psi_{\bT} ) +  \mathcal{O}\left( \delta^2 \right) .
\end{align}
Equation \eqref{Eq:DivQFIM} is derived in Appendix \ref{App:OptPSQFIM} and requires that $M \delta^2 \ll t^2$.  $\hat{K} $ is independent of $i,j$, such that our distillation technique amplifies all nonzero entries of $\mathcal{I} (\bT | \psi^{\mathrm{ps}}_{\bT} ) $ simultaneously: $\mathcal{I} (\bT | \psi^{\mathrm{ps}}_{\bT} ) =     \mathcal{I} (\bT | \psi_{\bT} ) / t^2 +  \mathcal{O}\left( \delta^2 \right)$. Combining this result with Ineqs. \eqref{Eq:ScalQCBPure}, $ s^{(\mathrm{MI})}(\psi^{\mathrm{ps}}_{\bT} , W) \rightarrow 0$ when $M \delta^2 \ll t^2 \rightarrow 0$.\footnote{We have assumed that the nonpostslected quantum Fisher information  $\mathcal{I} (\bT | \psi_{\bT} )$ is nonsingular. If it is singular, one has to remove the singularity-producing parameters from the analysis.}   Finally, the probability of successful postselection is given by $\pps = t^{-2} + \mathcal{O}(\delta^2)$ (see App. \ref{App:OptPSQFIM}). Consequently, the distillation of information is \textit{lossless}: $\pps \times \mathcal{I} (\bT | \psi^{\mathrm{ps}}_{\bT} )    = \mathcal{I} (\bT | \psi_{\bT} ) + \mathcal{O}(\delta^2)$.\footnote{Appendix \ref{App:PSQuant} shows that the geometric quantumness $\mathcal{Q}$ in Ineqs. \eqref{Eq:QuantScal} is constant [to $\mathcal{O}(\delta^2)$] with respect to the postselection.}  This concludes our constructive proof.$\square$

\section{Applications}

By distilling the multiparameter Fisher information, the intensity of output states is reduced. This can lead to  learnability improvements by allowing metrologists and machine learners  to use an intensity of input states that normally would have caused the output detectors to saturate. The information content available in the distilled, low-intensity output is identical to what the nondistilled, high-intensity output would have been. 

As an example, consider encoding an image in quantum states $\rT$.  $\bT=(\theta_1,...,\theta_M)$  is a vector of the image's pixels' intensities.  Perhaps our task is to find imperfections in the image-encoding procedure of a certain image $\bT^{\star}$. Then $\hat{\rho}_{\bT^{\star}}$ is a good initial guess to learn the imperfectly encoded, true image $\bT \approx \bT^{\star}$. Or perhaps we want to learn an image that  deviates slightly from a blank image. Then $\bT^0 = \bm{0}$ and $\rTe = \hat{\rho}_0$ is a good initial guess. Our distillation protocol allows us to both avoid detector saturation and increase sensitivity, without losing information, when measuring $\rT$ to learn the image. 

A particle-number detector will suffer from a \textit{dead time}, the time needed to reset the detector after triggering it. In the jargon of experimental costs: The dead time associates a temporal cost with the measurement \cite{Lupu20}. Also, measurements call for postprocessing, which cost further time and computation. 
Under the right conditions, our distillation protocol enables an experimentalist  to incur the final-measurement's cost only when the probe state carries a great deal of information. The ``right conditions'' are when the postselection is experimentally  \textit{cheaper} than the final measurement. 

Many quantum schemes  can be sped up by using several quantum processors in parallel \cite{Tang19, Grimsley19, Yordanov20}. By using our protocol to distill the output from  parallel processors, it could be possible to reduce the number of final-measurement apparatuses in  setups, decreasing the monetary cost of  parallel-processor schemes.

One can also envision scenarios where the encoding and final measurements are  spatially separated and connected by quantum channels. Our distillation protocol allows the rate of quantum-state transmission to decrease, whilst keeping the average information flow constant.

\section{Conclusion}
The quantum Fisher information matrix enables scalar quantification of quantum learnability in  multiparameter metrology and machine learning. We have shown that there exist upper and lower classical bounds on the entries in the quantum Fisher information matrix. Kirkwood-Dirac negativity, a narrower nonclassicality concept than noncommutation, allows the entries to break these bounds (Thm. \ref{Th:NCQFIM}). Motivated by this result, we designed a  protocol that uses a quantum analogue of Bayes' theorem to amplify uniformly  the nonzero entries in the  quantum Fisher information matrix.  This translates into the ability to probabilistically distill quantum learnability in a lossless fashion. We proved (Thm. \ref{Th:DivPSLearn})  that there is no upper bound on how much multiparameter information can be distilled into a small number of states. From a theoretical perspective, our results shed new light on the quantum Fisher information matrix and  generalizes, to the multiparameter-quantum-learnability regime, previous results in single-parameter postselected metrology and weak-value amplification. From a practical perspective, our results could mitigate the impact of detector imperfections and enable simplified setups in parallelized quantum schemes.


\textit{Acknowledgements.---}The authors would like to thank Crispin Barnes, Rafal Demkowicz-Dobrzanski, Bobak Kiani, Aleks Lasek, Zi-Wen Liu, Seth Lloyd, Noah Lupu Gladstein, Milad Marvian, Yordan Yordanov, and Nicole Yunger Halpern for useful discussions. This work was supported by the EPSRC, Lars Hierta's Memorial Foundation, and Girton College.

\onecolumngrid

\newpage
\clearpage

 \appendix

 \section{Derivation of Eq. \eqref{Eq:PSqfim}}
 \label{App:PSQFIM}

This appendix derives Eq. \eqref{Eq:PSqfim}. We proceed by changing the quantum state in 
\begin{equation}
	\mathcal{I}_{i,j}(\bT | \psi_{\bT} ) = 4\Re \left[ \braket{\partial_i\psi_{\bT}|\partial_j\psi_{\bT}} - \braket{\partial_i\psi_{\bT}|\psi_{\bT}}\braket{\psi_{\bT}|\partial_j\psi_{\bT}} \right],
\end{equation}
to  $\ket{\psi_{\bT}} \rightarrow  \ket{\psi_{\bT}^{\mathrm{ps}}} \equiv  \hat{K} \ket{\psi_{\bT}} / \sqrt{\pps}$. Remember that $\hat{F} = \hat{K}^{\dagger}\hat{K}$.

The first inner product is given by
\begin{align}
\left( \partial_i \frac{ \bra{\psi_{\bT}}  \hat{K}^{\dagger }}{\sqrt{\pps}} \right) \cdot \left( \partial_j \frac{\hat{K} \ket{\psi_{\bT}}  }{\sqrt{\pps}} \right) = & \left( \frac{ \bra{\partial_i  \psi_{\bT}}  \hat{K}^{\dagger }}{\sqrt{\pps}} - \frac{1}{2} \frac{ \bra{\psi_{\bT}}  \hat{K}^{\dagger }}{\left( \pps \right)^{\frac{3}{2}}} \partial_i \pps \right) \cdot \left(  \frac{\hat{K} \ket{ \partial_j \psi_{\bT}}  }{\sqrt{\pps}} - \frac{1}{2}  \frac{\hat{K} \ket{\psi_{\bT}}  }{\left( \pps \right)^{\frac{3}{2}}} \partial_j \pps \right) \\
= &  \frac{ \bra{\partial_i  \psi_{\bT}}  \hat{F} \ket{ \partial_j \psi_{\bT}} }{\pps} - \frac{1}{2} \frac{ \bra{\psi_{\bT}}   \hat{F} \ket{ \partial_j \psi_{\bT}} }{\left( \pps \right)^{2}} \partial_i \pps   \nonumber \\
& - \frac{1}{2}  \frac{ \bra{\partial_i  \psi_{\bT}}  \hat{F} \ket{\psi_{\bT}}  }{\left( \pps \right)^{2}} \partial_j \pps   + \frac{1}{4}  \frac{ \bra{\psi_{\bT}} \hat{F} \ket{\psi_{\bT}}  }{\left( \pps \right)^{3}} \left( \partial_i \pps \right) \left(  \partial_j \pps  \right) 
\\
= &  \frac{ \bra{\partial_i  \psi_{\bT}}  \hat{F} \ket{ \partial_j \psi_{\bT}} }{\pps} - \frac{1}{2} \frac{ \bra{\psi_{\bT}}   \hat{F} \ket{ \partial_j \psi_{\bT}} }{\left( \pps \right)^{2}} \partial_i \pps   \nonumber \\
& - \frac{1}{2}  \frac{ \bra{\partial_i  \psi_{\bT}}  \hat{F} \ket{\psi_{\bT}}  }{\left( \pps \right)^{2}} \partial_j \pps   + \frac{1}{4}  \frac{ \left( \partial_i \pps \right) \left(  \partial_j \pps  \right)  }{\left( \pps \right)^{2}} .
\end{align}
The last equality follows from $\pps =  \bra{\psi_{\bT}} \hat{F} \ket{\psi_{\bT}} $.

The second inner product is given by
\begin{align}
\left( \partial_i \frac{ \bra{\psi_{\bT}}  \hat{K}^{\dagger }}{\sqrt{\pps}} \right) \cdot \left( \frac{\hat{K} \ket{\psi_{\bT}}  }{\sqrt{\pps}} \right) = &  \frac{ \bra{\partial_i  \psi_{\bT}}  \hat{F}  \ket{\psi_{\bT}} }{\pps} - \frac{1}{2}  \frac{ \bra{  \psi_{\bT}}  \hat{F}  \ket{\psi_{\bT}} }{\left( \pps\right)^{2}} \left( \partial_i \pps\right) =   \frac{ \bra{\partial_i  \psi_{\bT}}  \hat{F}  \ket{\psi_{\bT}} }{\pps} - \frac{1}{2}  \frac{ \left( \partial_i \pps\right) }{ \pps} .
\end{align}

The third inner product is given by
\begin{align}
\left( \frac{ \bra{\psi_{\bT}}  \hat{K}^{\dagger }}{\sqrt{\pps}} \right) \cdot \left( \partial_j  \frac{\hat{K} \ket{\psi_{\bT}}  }{\sqrt{\pps}} \right) = &  \frac{ \bra{  \psi_{\bT}}  \hat{F}  \ket{\partial_j \psi_{\bT}} }{\pps} - \frac{1}{2}  \frac{ \bra{  \psi_{\bT}}  \hat{F}  \ket{\psi_{\bT}} }{\left( \pps\right)^{2}} \left( \partial_j \pps\right) =  \frac{ \bra{  \psi_{\bT}}  \hat{F}  \ket{\partial_j \psi_{\bT}} }{\pps} -  \frac{1}{2}  \frac{ \left( \partial_j \pps\right) }{ \pps} .
\end{align}

Combining these expressions:
\begin{equation}
	\mathcal{I}_{i,j}(\bT | \psi^{\mathrm{ps}}_{\bT}  ) = 4\Re \left[ \frac{1}{\pps} \braket{\partial_i\psi_{\bT} |  \hat{F} |  \partial_j\psi_{\bT}} - \frac{1}{\left( \pps \right)^2} \braket{\partial_i\psi_{\bT}| \hat{F} | \psi_{\bT}}\braket{\psi_{\bT}| \hat{F} | \partial_j\psi_{\bT}} \right].
\end{equation}
 Transforming the inner products to traces, we retrieve Eq. \eqref{Eq:PSqfim}:
 \begin{align}
	\mathcal{I}_{i,j} (\bT | \psi^{\mathrm{ps}}_{\bT} ) = & 4\Re \bigg\{ \frac{1}{\pps} \Tr\left[ \hat{F} \hat{\tilde{A}}_j \rT \hat{\tilde{A}}_i \right]  - \frac{1}{(\pps)^2} \Tr\left[\hat{F} \rT \hat{\tilde{A}}_i \right]\Tr\left[ \hat{F} \hat{\tilde{A}}_j \rT \right] \bigg\} .
\end{align}
 Here, we have used that
 \begin{align}
 \ket{\partial_j\psi_{\bT}} = &  \partial_j \hat{U}(\bT) \ket{\psi_0} \\
 = & \hat{U}_M(\theta_M) \cdots \hat{U}_{j+1}(\theta_{j+1}) \hat{A}_j  \hat{U}_{j}(\theta_j) \cdots  \hat{U}_1(\theta_1) \ket{\psi_0} \\
 = &  \hat{U}_M(\theta_M) \cdots  \hat{U}_{j+1}(\theta_{j+1}) \hat{A}_j  \hat{U}_{j+1}^\dag(\theta_{j+1}) \cdots  \hat{U}_M^\dag(\theta_M) \hat{U}_M(\theta_M) \cdots  \hat{U}_{j+1}(\theta_{j+1})  \hat{U}_{j}(\theta_j) \cdots  \hat{U}_1(\theta_1) \ket{\psi_0}
 \\
 = &  \hat{\tilde{A}}_j \hat{U}_M(\theta_M) \cdots  \hat{U}_1(\theta_1) \ket{\psi_0}
  \\
 = &  \hat{\tilde{A}}_j \hat{U}(\bT) \ket{\psi_0}
   \\
 = &  \hat{\tilde{A}}_j  \ket{\psi_{\bT}} ,
 \end{align}
where $\hat{\tilde{A}}_j \equiv  \hat{U}_M(\theta_M) \cdots  \hat{U}_{j+1}(\theta_{j+1}) \hat{A}_j  \hat{U}_{j+1}^\dag(\theta_{j+1}) \cdots  \hat{U}_M^\dag(\theta_M)$ for $j < M$. For $j=M$, $\hat{\tilde{A}}_M = \hat{A}_M$.

 \section{Derivation of Eq. \eqref{Eq:DivQFIM}}
 \label{App:OptPSQFIM}
 
 This appendix evaluates the postselected quantum Fisher information matrix $ \mathcal{I}_{i,j} (\bT | \psi^{\mathrm{ps}}_{\bT} ) $ [Eq. \eqref{Eq:PSqfim}]  for the choice of Kraus operator presented in Eq. \eqref{Eq:OptF}: $\hat{K} = (t-1) \rTe +  \hat{1}  $. This Kraus operator generates the postselection operator $\hat{F} =  (t^2 - 1) \rTe +  \hat{1}  $. In the main text, we defined $\rTe \equiv \hat{U}(\bTe) \hat{\rho}_0 \hat{U}^{\dagger}(\bTe) = \rT +  \left[ \rT \, , \, \hat{D} \right] +  \mathcal{O}\left( \delta^2 \right)$. The following calculations assume that $M \delta^2 \ll t^2$.  We define $\hat{C} \equiv [\rT \, , \, \hat{D} ]$ to simplify notation. We begin by evaluating the individual terms of  Eq. \eqref{Eq:PSqfim}. Then, we combine these terms. 
 
 First, we calculate the postselection probability $\pps$ in Eq. \eqref{Eq:PSqfim}:
 \begin{align}
 \pps = &    \Tr\left[ \hat{F}  \rT  \right] \\
 = &   \Tr\left[ \left(  \rTe  (t^2 -1 )+ \hat{1}  \right)  \rT  \right] \\
 = & (t^2 -1 ) \Tr\left[  \rTe   \rT  \right] + 1 \\
 = & (t^2 -1 ) \Tr\left[  \left( \rT +  \hat{C} \right)   \rT  \right] + 1  +  \mathcal{O}\left( \delta^2 \right) \\
  = & (t^2 -1 ) \Tr\left[   \rT   \right] + 1  +  \mathcal{O}\left( \delta^2 \right)  \\
  = & t^2   +  \mathcal{O}\left( \delta^2 \right) .
 \end{align}
Here, we have used that $ \Tr[   \rT \hat{C} ] = 0 $.

 Second, we calculate the first trace in Eq. \eqref{Eq:PSqfim}:
 \begin{align}
 \Tr\left[ \hat{F} \hat{\tilde{A}}_j \rT \hat{\tilde{A}}_i \right]   & =    \Tr\left[ \left(  \rTe  (t^2 -1 )+ \hat{1}  \right) \hat{\tilde{A}}_j \rT \hat{\tilde{A}}_i \right]  \\
	 & =    \Tr\left[ \left(    (t^2 -1 ) (\rT + \hat{C} )+ \hat{1}  \right) \hat{\tilde{A}}_j \rT \hat{\tilde{A}}_i \right] + \mathcal{O}\left( \delta \right)  \\
	 & =  (t^2 -1 )   \Tr\left[     \hat{C}    \hat{\tilde{A}}_j \rT \hat{\tilde{A}}_i \right] + (t^2 -1 )\Tr\left[     \rT \hat{\tilde{A}}_i \right] \Tr\left[    \hat{\tilde{A}}_j \rT \right] +
	 \Tr\left[ \hat{\tilde{A}}_j \rT \hat{\tilde{A}}_i \right] + \mathcal{O}\left( \delta \right) .
\end{align}
 
  Third, we calculate the second trace in Eq. \eqref{Eq:PSqfim}:
\begin{align}
	 \Tr\left[\hat{F} \rT \hat{\tilde{A}}_i \right] 	 = &   \Tr\left[ \left(  \rTe  (t^2 -1 )+ \hat{1}  \right)  \rT \hat{\tilde{A}}_i \right]]  \\
	 = &  \Tr\left[ \left(    (t^2 -1 ) (\rT + \hat{C} )+ \hat{1}  \right)  \rT \hat{\tilde{A}}_i \right]
	 + \mathcal{O}\left( \delta^2 \right)  \\
	 = &  \Tr\left[ \rT \hat{\tilde{A}}_i \right]
	 +   \Tr\left[    (t^2 -1 ) (\rT + \hat{C} ) \rT \hat{\tilde{A}}_i \right]
	 + \mathcal{O}\left( \delta^2 \right)  \\
	 	 = &  \Tr\left[ \rT \hat{\tilde{A}}_i \right]
	 +   (t^2 -1 ) \Tr\left[     \rT \hat{\tilde{A}}_i \right]
	 +  (t^2 -1 ) \Tr\left[     \hat{C} \rT \hat{\tilde{A}}_i \right]
	 + \mathcal{O}\left( \delta^2 \right)  \\
	 	 	 = & t^2  \Tr\left[ \rT \hat{\tilde{A}}_i \right]
	 +  (t^2 -1 ) \Tr\left[     \hat{C} \rT \hat{\tilde{A}}_i \right]
	 + \mathcal{O}\left( \delta^2 \right)  .
\end{align} 

Fourth,  we calculate the third trace in Eq. \eqref{Eq:PSqfim}:
\begin{align}
	 \Tr\left[\hat{F} \hat{\tilde{A}}_j \rT  \right] 	 = &   \Tr\left[ \left(  \rTe  (t^2 -1 )+ \hat{1}  \right) \hat{\tilde{A}}_j \rT  \right]]  \\
	 = &  \Tr\left[ \left(    (t^2 -1 ) (\rT + \hat{C} )+ \hat{1}  \right) \hat{\tilde{A}}_j  \rT \right]
	 + \mathcal{O}\left( \delta^2 \right)  \\
	 = &  \Tr\left[ \hat{\tilde{A}}_j \rT  \right]
	 +   \Tr\left[    (t^2 -1 ) (\rT + \hat{C} ) \hat{\tilde{A}}_j  \rT  \right]
	 + \mathcal{O}\left( \delta^2 \right)  \\
	 	 = &  \Tr\left[ \hat{\tilde{A}}_j \rT  \right]
	 +   (t^2 -1 ) \Tr\left[   \hat{\tilde{A}}_j  \rT  \right]
	 +  (t^2 -1 ) \Tr\left[     \hat{C} \hat{\tilde{A}}_j \rT  \right]
	 + \mathcal{O}\left( \delta^2 \right)  \\
	 	 	 = & t^2  \Tr\left[ \hat{\tilde{A}}_j \rT  \right]
	 +  (t^2 -1 ) \Tr\left[     \hat{C} \hat{\tilde{A}}_j  \rT  \right]
	 + \mathcal{O}\left( \delta^2 \right)  .
\end{align} 
 
Fifth, we calculate the product of the second and third trace in Eq. \eqref{Eq:PSqfim}:
 \begin{align}
 \Tr\left[\hat{F} \rT \hat{\tilde{A}}_i \right] 	 \times 	 \Tr\left[\hat{F} \hat{\tilde{A}}_j \rT  \right] = &  t^4\Tr\left[ \rT \hat{\tilde{A}}_i \right]  \Tr\left[ \hat{\tilde{A}}_j \rT  \right] \nonumber \\
	 & + t^2 (t^2 -1 ) \left\{    \Tr\left[ \rT \hat{\tilde{A}}_i \right] \Tr\left[     \hat{C} \hat{\tilde{A}}_j  \rT  \right] +  \Tr\left[ \hat{\tilde{A}}_j \rT  \right]  \Tr\left[     \hat{C} \rT \hat{\tilde{A}}_i \right]  \right\} + \mathcal{O}\left( \delta^2 \right) \\
	 = &  t^4\Tr\left[ \rT \hat{\tilde{A}}_i \right]  \Tr\left[ \hat{\tilde{A}}_j \rT  \right] \nonumber \\
	 & + t^2 (t^2 -1 ) \Big\{    \Tr\left[ \rT \hat{\tilde{A}}_i \right] \Tr\left[    \rT  \hat{D} \hat{\tilde{A}}_j  \rT  \right] - \Tr\left[ \rT \hat{\tilde{A}}_i \right] \Tr\left[     \hat{D} \rT \hat{\tilde{A}}_j  \rT  \right] \nonumber \\
	 & +
	  \Tr\left[ \hat{\tilde{A}}_j \rT  \right]  \Tr\left[   \rT  \hat{D} \rT \hat{\tilde{A}}_i \right] - \Tr\left[ \hat{\tilde{A}}_j \rT  \right]  \Tr\left[     \hat{D} \rT \rT \hat{\tilde{A}}_i \right]  \Big\} + \mathcal{O}\left( \delta^2 \right) \\
	 = &  t^4\Tr\left[ \rT \hat{\tilde{A}}_i \right]  \Tr\left[ \hat{\tilde{A}}_j \rT  \right] \nonumber \\
	 & + t^2 (t^2 -1 ) \Big\{    \Tr\left[ \rT \hat{\tilde{A}}_i \right] \Tr\left[    \rT  \hat{D} \hat{\tilde{A}}_j   \right] - \Tr\left[ \hat{\tilde{A}}_j \rT  \right]  \Tr\left[     \hat{D} \rT \hat{\tilde{A}}_i \right]  \Big\} + \mathcal{O}\left( \delta^2 \right) \\
	 = &  t^4\Tr\left[ \rT \hat{\tilde{A}}_i \right]  \Tr\left[ \hat{\tilde{A}}_j \rT  \right] + t^2 (t^2 -1 )    \Tr\left[  \left(  \rT \hat{D} -\hat{D} \rT \right)  \hat{\tilde{A}}_j \rT  \hat{\tilde{A}}_i \right]  + \mathcal{O}\left( \delta^2 \right) \\
	 = &  t^4\Tr\left[ \rT \hat{\tilde{A}}_i \right]  \Tr\left[ \hat{\tilde{A}}_j \rT  \right] + t^2 (t^2 -1 )     \Tr\left[     \hat{C}    \hat{\tilde{A}}_j \rT \hat{\tilde{A}}_i \right]   + \mathcal{O}\left( \delta^2 \right) .
\end{align}  
Again, we have used the result that $ \Tr[   \rT \hat{C} ] = 0 $.

Finally, we combine the calculated expressions:
  \begin{align}
\mathcal{I}_{i,j} (\bT | \psi^{\mathrm{ps}}_{\bT} ) 
	= &  4 \Re \Bigg\{ \frac{(t^2 -1 )}{t^2}   \Tr\left[     \hat{C}    \hat{\tilde{A}}_j \rT \hat{\tilde{A}}_i \right] +\frac{(t^2 -1 )}{t^2} \Tr\left[     \rT \hat{\tilde{A}}_i \right] \Tr\left[    \hat{\tilde{A}}_j \rT \right] + \frac{1}{t^2}
	 \Tr\left[ \hat{\tilde{A}}_j \rT \hat{\tilde{A}}_i \right] \nonumber \\
	 & -  \Tr\left[ \rT \hat{\tilde{A}}_i \right]  \Tr\left[ \hat{\tilde{A}}_j \rT  \right] - \frac{(t^2 -1 )}{t^2}     \Tr\left[     \hat{C}    \hat{\tilde{A}}_j \rT \hat{\tilde{A}}_i \right]  \Bigg\}  + \mathcal{O}\left( \delta^2 \right) \\
	 = & \frac{4}{t^2}\Re \bigg\{  \Tr\left[ \hat{\tilde{A}}_j \rT \hat{\tilde{A}}_i \right] -  \Tr\left[\rT \hat{\tilde{A}}_i \right]\Tr\left[  \hat{\tilde{A}}_j \rT \right] \bigg\} \\ 
	 = &  \frac{ 1}{ t^2  } \mathcal{I}_{i,j} (\bT | \psi_{\bT} ) +  \mathcal{O}\left( \delta^2 \right) .
\end{align}
 This is the expression given in Eq. \eqref{Eq:DivQFIM}.

  \section{Postselected geometric quantumness}
 \label{App:PSQuant}

The geometric quantumness measure $\mathcal{Q} $ in Ineqs. \eqref{Eq:QuantScal} is given by 
\begin{equation}
\mathcal{Q} = || i \mathcal{I}^{-1} (\bT | \psi_{\bT} ) \mathcal{J}(\bT | \psi_{\bT} )  ||_{\infty} ,
\end{equation}
where $|| X ||_{\infty}$ denotes the largest eigenvalue of $X$. $ \mathcal{J}(\bT | \psi_{\bT} )$ is the Uhlmann curvature\footnote{For pure states, $ \mathcal{J}(\bT | \psi_{\bT} )$ is (four times) the imaginary part of the  quantum geometric tensor. $ \mathcal{I}(\bT | \psi_{\bT} )$ is (four times) the real part. } \cite{Carollo18} given by
\begin{equation}
	\mathcal{J}_{i,j}(\bT | \psi_{\bT} ) = 4\Im \left[ \braket{\partial_i\psi_{\bT}|\partial_j\psi_{\bT}} - \braket{\partial_i\psi_{\bT}|\psi_{\bT}}\braket{\psi_{\bT}|\partial_j\psi_{\bT}} \right]  .
\end{equation}

The same tricks used in Appendix \ref{App:OptPSQFIM} can be used to show that 
\begin{equation}
\mathcal{J}_{i,j}(\bT | \psi^{\mathrm{ps}}_{\bT} ) = \frac{ 1}{ t^2  } \mathcal{J}_{i,j} (\bT | \psi_{\bT} ) +  \mathcal{O}\left( \delta^2 \right) .
\end{equation}
Thus, at least to $\mathcal{O}(\delta^2)$, the geometric quantumness $\mathcal{Q}$ is constant with respect to the postselection.

\bibliography{MultiPSMet}

\begin{thebibliography}{75}%
\makeatletter
\providecommand \@ifxundefined [1]{%
 \@ifx{#1\undefined}
}%
\providecommand \@ifnum [1]{%
 \ifnum #1\expandafter \@firstoftwo
 \else \expandafter \@secondoftwo
 \fi
}%
\providecommand \@ifx [1]{%
 \ifx #1\expandafter \@firstoftwo
 \else \expandafter \@secondoftwo
 \fi
}%
\providecommand \natexlab [1]{#1}%
\providecommand \enquote  [1]{``#1''}%
\providecommand \bibnamefont  [1]{#1}%
\providecommand \bibfnamefont [1]{#1}%
\providecommand \citenamefont [1]{#1}%
\providecommand \href@noop [0]{\@secondoftwo}%
\providecommand \href [0]{\begingroup \@sanitize@url \@href}%
\providecommand \@href[1]{\@@startlink{#1}\@@href}%
\providecommand \@@href[1]{\endgroup#1\@@endlink}%
\providecommand \@sanitize@url [0]{\catcode `\\12\catcode `\$12\catcode
  `\&12\catcode `\#12\catcode `\^12\catcode `\_12\catcode `\%12\relax}%
\providecommand \@@startlink[1]{}%
\providecommand \@@endlink[0]{}%
\providecommand \url  [0]{\begingroup\@sanitize@url \@url }%
\providecommand \@url [1]{\endgroup\@href {#1}{\urlprefix }}%
\providecommand \urlprefix  [0]{URL }%
\providecommand \Eprint [0]{\href }%
\providecommand \doibase [0]{http://dx.doi.org/}%
\providecommand \selectlanguage [0]{\@gobble}%
\providecommand \bibinfo  [0]{\@secondoftwo}%
\providecommand \bibfield  [0]{\@secondoftwo}%
\providecommand \translation [1]{[#1]}%
\providecommand \BibitemOpen [0]{}%
\providecommand \bibitemStop [0]{}%
\providecommand \bibitemNoStop [0]{.\EOS\space}%
\providecommand \EOS [0]{\spacefactor3000\relax}%
\providecommand \BibitemShut  [1]{\csname bibitem#1\endcsname}%
\let\auto@bib@innerbib\@empty
\bibitem [{\citenamefont {Giovannetti}\ \emph {et~al.}(2011)\citenamefont
  {Giovannetti}, \citenamefont {Lloyd},\ and\ \citenamefont
  {Maccone}}]{Giovanetti11}%
  \BibitemOpen
  \bibfield  {author} {\bibinfo {author} {\bibfnamefont {V.}~\bibnamefont
  {Giovannetti}}, \bibinfo {author} {\bibfnamefont {S.}~\bibnamefont {Lloyd}},
  \ and\ \bibinfo {author} {\bibfnamefont {L.}~\bibnamefont {Maccone}},\
  }\href@noop {} {\bibfield  {journal} {\bibinfo  {journal} {Nature photonics}\
  }\textbf {\bibinfo {volume} {5}},\ \bibinfo {pages} {222} (\bibinfo {year}
  {2011})}\BibitemShut {NoStop}%
\bibitem [{\citenamefont {Maccone}(2013)}]{Maccone13}%
  \BibitemOpen
  \bibfield  {author} {\bibinfo {author} {\bibfnamefont {L.}~\bibnamefont
  {Maccone}},\ }\href {\doibase 10.1103/PhysRevA.88.042109} {\bibfield
  {journal} {\bibinfo  {journal} {Phys. Rev. A}\ }\textbf {\bibinfo {volume}
  {88}},\ \bibinfo {pages} {042109} (\bibinfo {year} {2013})}\BibitemShut
  {NoStop}%
\bibitem [{\citenamefont {Szczykulska}\ \emph {et~al.}(2016)\citenamefont
  {Szczykulska}, \citenamefont {Baumgratz},\ and\ \citenamefont
  {Datta}}]{Szczykulska16}%
  \BibitemOpen
  \bibfield  {author} {\bibinfo {author} {\bibfnamefont {M.}~\bibnamefont
  {Szczykulska}}, \bibinfo {author} {\bibfnamefont {T.}~\bibnamefont
  {Baumgratz}}, \ and\ \bibinfo {author} {\bibfnamefont {A.}~\bibnamefont
  {Datta}},\ }\href {\doibase 10.1080/23746149.2016.1230476} {\bibfield
  {journal} {\bibinfo  {journal} {Advances in Physics: X}\ }\textbf {\bibinfo
  {volume} {1}},\ \bibinfo {pages} {621} (\bibinfo {year} {2016})},\ \Eprint
  {http://arxiv.org/abs/https://doi.org/10.1080/23746149.2016.1230476}
  {https://doi.org/10.1080/23746149.2016.1230476} \BibitemShut {NoStop}%
\bibitem [{\citenamefont {Kiani}\ \emph {et~al.}(2021)\citenamefont {Kiani},
  \citenamefont {De~Palma}, \citenamefont {Marvian}, \citenamefont {Liu},\ and\
  \citenamefont {Lloyd}}]{Kiani21}%
  \BibitemOpen
  \bibfield  {author} {\bibinfo {author} {\bibfnamefont {B.~T.}\ \bibnamefont
  {Kiani}}, \bibinfo {author} {\bibfnamefont {G.}~\bibnamefont {De~Palma}},
  \bibinfo {author} {\bibfnamefont {M.}~\bibnamefont {Marvian}}, \bibinfo
  {author} {\bibfnamefont {Z.-W.}\ \bibnamefont {Liu}}, \ and\ \bibinfo
  {author} {\bibfnamefont {S.}~\bibnamefont {Lloyd}},\ }\href
  {https://arxiv.org/abs/2101.03037} {\bibfield  {journal} {\bibinfo  {journal}
  {arXiv preprint arXiv:2101.03037}\ } (\bibinfo {year} {2021})}\BibitemShut
  {NoStop}%
\bibitem [{\citenamefont {Arvidsson-Shukur}\ \emph
  {et~al.}(2020{\natexlab{a}})\citenamefont {Arvidsson-Shukur}, \citenamefont
  {Drori},\ and\ \citenamefont {Halpern}}]{ArvShuk20}%
  \BibitemOpen
  \bibfield  {author} {\bibinfo {author} {\bibfnamefont {D.~R.}\ \bibnamefont
  {Arvidsson-Shukur}}, \bibinfo {author} {\bibfnamefont {J.~C.}\ \bibnamefont
  {Drori}}, \ and\ \bibinfo {author} {\bibfnamefont {N.~Y.}\ \bibnamefont
  {Halpern}},\ }\href {\doibase https://arxiv.org/abs/2009.04468} {\bibfield
  {journal} {\bibinfo  {journal} {arXiv preprint arXiv:2009.04468}\ } (\bibinfo
  {year} {2020}{\natexlab{a}}),\ https://arxiv.org/abs/2009.04468}\BibitemShut
  {NoStop}%
\bibitem [{\citenamefont {Braunstein}\ and\ \citenamefont
  {Caves}(1994)}]{Braunstein94}%
  \BibitemOpen
  \bibfield  {author} {\bibinfo {author} {\bibfnamefont {S.~L.}\ \bibnamefont
  {Braunstein}}\ and\ \bibinfo {author} {\bibfnamefont {C.~M.}\ \bibnamefont
  {Caves}},\ }\href {\doibase 10.1103/PhysRevLett.72.3439} {\bibfield
  {journal} {\bibinfo  {journal} {Phys. Rev. Lett.}\ }\textbf {\bibinfo
  {volume} {72}},\ \bibinfo {pages} {3439} (\bibinfo {year}
  {1994})}\BibitemShut {NoStop}%
\bibitem [{\citenamefont {Liu}\ \emph {et~al.}(2019)\citenamefont {Liu},
  \citenamefont {Yuan}, \citenamefont {Lu},\ and\ \citenamefont
  {Wang}}]{Liu19}%
  \BibitemOpen
  \bibfield  {author} {\bibinfo {author} {\bibfnamefont {J.}~\bibnamefont
  {Liu}}, \bibinfo {author} {\bibfnamefont {H.}~\bibnamefont {Yuan}}, \bibinfo
  {author} {\bibfnamefont {X.-M.}\ \bibnamefont {Lu}}, \ and\ \bibinfo {author}
  {\bibfnamefont {X.}~\bibnamefont {Wang}},\ }\href
  {https://iopscience.iop.org/article/10.1088/1751-8121/ab5d4d/meta} {\bibfield
   {journal} {\bibinfo  {journal} {Journal of Physics A: Mathematical and
  Theoretical}\ }\textbf {\bibinfo {volume} {53}},\ \bibinfo {pages} {023001}
  (\bibinfo {year} {2019})}\BibitemShut {NoStop}%
\bibitem [{\citenamefont {Albarelli}\ \emph {et~al.}(2020)\citenamefont
  {Albarelli}, \citenamefont {Barbieri}, \citenamefont {Genoni},\ and\
  \citenamefont {Gianani}}]{Albarelli20}%
  \BibitemOpen
  \bibfield  {author} {\bibinfo {author} {\bibfnamefont {F.}~\bibnamefont
  {Albarelli}}, \bibinfo {author} {\bibfnamefont {M.}~\bibnamefont {Barbieri}},
  \bibinfo {author} {\bibfnamefont {M.}~\bibnamefont {Genoni}}, \ and\ \bibinfo
  {author} {\bibfnamefont {I.}~\bibnamefont {Gianani}},\ }\href {\doibase
  https://doi.org/10.1016/j.physleta.2020.126311} {\bibfield  {journal}
  {\bibinfo  {journal} {Physics Letters A}\ }\textbf {\bibinfo {volume}
  {384}},\ \bibinfo {pages} {126311} (\bibinfo {year} {2020})}\BibitemShut
  {NoStop}%
\bibitem [{\citenamefont {Rao}(1992)}]{Rao92}%
  \BibitemOpen
  \bibfield  {author} {\bibinfo {author} {\bibfnamefont {C.~R.}\ \bibnamefont
  {Rao}},\ }in\ \href@noop {} {\emph {\bibinfo {booktitle} {Breakthroughs in
  statistics}}}\ (\bibinfo  {publisher} {Springer},\ \bibinfo {year} {1992})\
  pp.\ \bibinfo {pages} {235--247}\BibitemShut {NoStop}%
\bibitem [{\citenamefont {Cram{\'e}r}(2016)}]{Cramer16}%
  \BibitemOpen
  \bibfield  {author} {\bibinfo {author} {\bibfnamefont {H.}~\bibnamefont
  {Cram{\'e}r}},\ }\href@noop {} {\emph {\bibinfo {title} {Mathematical methods
  of statistics (PMS-9)}}},\ Vol.~\bibinfo {volume} {9}\ (\bibinfo  {publisher}
  {Princeton University Press},\ \bibinfo {year} {2016})\BibitemShut {NoStop}%
\bibitem [{\citenamefont {Ballester}(2004)}]{Ballester04}%
  \BibitemOpen
  \bibfield  {author} {\bibinfo {author} {\bibfnamefont {M.~A.}\ \bibnamefont
  {Ballester}},\ }\href {\doibase 10.1103/PhysRevA.69.022303} {\bibfield
  {journal} {\bibinfo  {journal} {Phys. Rev. A}\ }\textbf {\bibinfo {volume}
  {69}},\ \bibinfo {pages} {022303} (\bibinfo {year} {2004})}\BibitemShut
  {NoStop}%
\bibitem [{\citenamefont {Imai}\ and\ \citenamefont
  {Fujiwara}(2007)}]{Fujiwara07}%
  \BibitemOpen
  \bibfield  {author} {\bibinfo {author} {\bibfnamefont {H.}~\bibnamefont
  {Imai}}\ and\ \bibinfo {author} {\bibfnamefont {A.}~\bibnamefont
  {Fujiwara}},\ }\href
  {https://iopscience.iop.org/article/10.1088/1751-8113/40/16/009} {\bibfield
  {journal} {\bibinfo  {journal} {Journal of Physics A: Mathematical and
  Theoretical}\ }\textbf {\bibinfo {volume} {40}},\ \bibinfo {pages} {4391}
  (\bibinfo {year} {2007})}\BibitemShut {NoStop}%
\bibitem [{\citenamefont {Genoni}\ \emph {et~al.}(2013)\citenamefont {Genoni},
  \citenamefont {Paris}, \citenamefont {Adesso}, \citenamefont {Nha},
  \citenamefont {Knight},\ and\ \citenamefont {Kim}}]{Genoni13}%
  \BibitemOpen
  \bibfield  {author} {\bibinfo {author} {\bibfnamefont {M.~G.}\ \bibnamefont
  {Genoni}}, \bibinfo {author} {\bibfnamefont {M.~G.~A.}\ \bibnamefont
  {Paris}}, \bibinfo {author} {\bibfnamefont {G.}~\bibnamefont {Adesso}},
  \bibinfo {author} {\bibfnamefont {H.}~\bibnamefont {Nha}}, \bibinfo {author}
  {\bibfnamefont {P.~L.}\ \bibnamefont {Knight}}, \ and\ \bibinfo {author}
  {\bibfnamefont {M.~S.}\ \bibnamefont {Kim}},\ }\href {\doibase
  10.1103/PhysRevA.87.012107} {\bibfield  {journal} {\bibinfo  {journal} {Phys.
  Rev. A}\ }\textbf {\bibinfo {volume} {87}},\ \bibinfo {pages} {012107}
  (\bibinfo {year} {2013})}\BibitemShut {NoStop}%
\bibitem [{\citenamefont {Humphreys}\ \emph {et~al.}(2013)\citenamefont
  {Humphreys}, \citenamefont {Barbieri}, \citenamefont {Datta},\ and\
  \citenamefont {Walmsley}}]{Humphreys13}%
  \BibitemOpen
  \bibfield  {author} {\bibinfo {author} {\bibfnamefont {P.~C.}\ \bibnamefont
  {Humphreys}}, \bibinfo {author} {\bibfnamefont {M.}~\bibnamefont {Barbieri}},
  \bibinfo {author} {\bibfnamefont {A.}~\bibnamefont {Datta}}, \ and\ \bibinfo
  {author} {\bibfnamefont {I.~A.}\ \bibnamefont {Walmsley}},\ }\href {\doibase
  10.1103/PhysRevLett.111.070403} {\bibfield  {journal} {\bibinfo  {journal}
  {Phys. Rev. Lett.}\ }\textbf {\bibinfo {volume} {111}},\ \bibinfo {pages}
  {070403} (\bibinfo {year} {2013})}\BibitemShut {NoStop}%
\bibitem [{\citenamefont {Pezz\`e}\ \emph {et~al.}(2017)\citenamefont
  {Pezz\`e}, \citenamefont {Ciampini}, \citenamefont {Spagnolo}, \citenamefont
  {Humphreys}, \citenamefont {Datta}, \citenamefont {Walmsley}, \citenamefont
  {Barbieri}, \citenamefont {Sciarrino},\ and\ \citenamefont
  {Smerzi}}]{Pezze17}%
  \BibitemOpen
  \bibfield  {author} {\bibinfo {author} {\bibfnamefont {L.}~\bibnamefont
  {Pezz\`e}}, \bibinfo {author} {\bibfnamefont {M.~A.}\ \bibnamefont
  {Ciampini}}, \bibinfo {author} {\bibfnamefont {N.}~\bibnamefont {Spagnolo}},
  \bibinfo {author} {\bibfnamefont {P.~C.}\ \bibnamefont {Humphreys}}, \bibinfo
  {author} {\bibfnamefont {A.}~\bibnamefont {Datta}}, \bibinfo {author}
  {\bibfnamefont {I.~A.}\ \bibnamefont {Walmsley}}, \bibinfo {author}
  {\bibfnamefont {M.}~\bibnamefont {Barbieri}}, \bibinfo {author}
  {\bibfnamefont {F.}~\bibnamefont {Sciarrino}}, \ and\ \bibinfo {author}
  {\bibfnamefont {A.}~\bibnamefont {Smerzi}},\ }\href {\doibase
  10.1103/PhysRevLett.119.130504} {\bibfield  {journal} {\bibinfo  {journal}
  {Phys. Rev. Lett.}\ }\textbf {\bibinfo {volume} {119}},\ \bibinfo {pages}
  {130504} (\bibinfo {year} {2017})}\BibitemShut {NoStop}%
\bibitem [{\citenamefont {Chen}\ and\ \citenamefont {Yuan}(2017)}]{Chen17}%
  \BibitemOpen
  \bibfield  {author} {\bibinfo {author} {\bibfnamefont {Y.}~\bibnamefont
  {Chen}}\ and\ \bibinfo {author} {\bibfnamefont {H.}~\bibnamefont {Yuan}},\
  }\href {https://iopscience.iop.org/article/10.1088/1367-2630/aa723d}
  {\bibfield  {journal} {\bibinfo  {journal} {New Journal of Physics}\ }\textbf
  {\bibinfo {volume} {19}},\ \bibinfo {pages} {063023} (\bibinfo {year}
  {2017})}\BibitemShut {NoStop}%
\bibitem [{\citenamefont {Liuzzo-Scorpo}\ \emph {et~al.}(2018)\citenamefont
  {Liuzzo-Scorpo}, \citenamefont {Correa}, \citenamefont {Pollock},
  \citenamefont {G{\'{o}}recka}, \citenamefont {Modi},\ and\ \citenamefont
  {Adesso}}]{Liuzzo18}%
  \BibitemOpen
  \bibfield  {author} {\bibinfo {author} {\bibfnamefont {P.}~\bibnamefont
  {Liuzzo-Scorpo}}, \bibinfo {author} {\bibfnamefont {L.~A.}\ \bibnamefont
  {Correa}}, \bibinfo {author} {\bibfnamefont {F.~A.}\ \bibnamefont {Pollock}},
  \bibinfo {author} {\bibfnamefont {A.}~\bibnamefont {G{\'{o}}recka}}, \bibinfo
  {author} {\bibfnamefont {K.}~\bibnamefont {Modi}}, \ and\ \bibinfo {author}
  {\bibfnamefont {G.}~\bibnamefont {Adesso}},\ }\href {\doibase
  10.1088/1367-2630/aac5b6} {\bibfield  {journal} {\bibinfo  {journal} {New
  Journal of Physics}\ }\textbf {\bibinfo {volume} {20}},\ \bibinfo {pages}
  {063009} (\bibinfo {year} {2018})}\BibitemShut {NoStop}%
\bibitem [{\citenamefont {Lipka-Bartosik}\ and\ \citenamefont
  {Demkowicz-Dobrza{\'{n}}ski}(2018)}]{Lipka18}%
  \BibitemOpen
  \bibfield  {author} {\bibinfo {author} {\bibfnamefont {P.}~\bibnamefont
  {Lipka-Bartosik}}\ and\ \bibinfo {author} {\bibfnamefont {R.}~\bibnamefont
  {Demkowicz-Dobrza{\'{n}}ski}},\ }\href {\doibase 10.1088/1751-8121/aae664}
  {\bibfield  {journal} {\bibinfo  {journal} {Journal of Physics A:
  Mathematical and Theoretical}\ }\textbf {\bibinfo {volume} {51}},\ \bibinfo
  {pages} {474001} (\bibinfo {year} {2018})}\BibitemShut {NoStop}%
\bibitem [{\citenamefont {Dressel}\ \emph {et~al.}(2014)\citenamefont
  {Dressel}, \citenamefont {Malik}, \citenamefont {Miatto}, \citenamefont
  {Jordan},\ and\ \citenamefont {Boyd}}]{Dressel14}%
  \BibitemOpen
  \bibfield  {author} {\bibinfo {author} {\bibfnamefont {J.}~\bibnamefont
  {Dressel}}, \bibinfo {author} {\bibfnamefont {M.}~\bibnamefont {Malik}},
  \bibinfo {author} {\bibfnamefont {F.~M.}\ \bibnamefont {Miatto}}, \bibinfo
  {author} {\bibfnamefont {A.~N.}\ \bibnamefont {Jordan}}, \ and\ \bibinfo
  {author} {\bibfnamefont {R.~W.}\ \bibnamefont {Boyd}},\ }\href {\doibase
  10.1103/RevModPhys.86.307} {\bibfield  {journal} {\bibinfo  {journal} {Rev.
  Mod. Phys.}\ }\textbf {\bibinfo {volume} {86}},\ \bibinfo {pages} {307}
  (\bibinfo {year} {2014})}\BibitemShut {NoStop}%
\bibitem [{\citenamefont {Harris}\ \emph {et~al.}(2017)\citenamefont {Harris},
  \citenamefont {Boyd},\ and\ \citenamefont {Lundeen}}]{Harris17}%
  \BibitemOpen
  \bibfield  {author} {\bibinfo {author} {\bibfnamefont {J.}~\bibnamefont
  {Harris}}, \bibinfo {author} {\bibfnamefont {R.~W.}\ \bibnamefont {Boyd}}, \
  and\ \bibinfo {author} {\bibfnamefont {J.~S.}\ \bibnamefont {Lundeen}},\
  }\href@noop {} {\bibfield  {journal} {\bibinfo  {journal} {Phys. Rev. Lett.}\
  }\textbf {\bibinfo {volume} {118}},\ \bibinfo {pages} {070802} (\bibinfo
  {year} {2017})}\BibitemShut {NoStop}%
\bibitem [{\citenamefont {Xu}\ \emph {et~al.}(2020)\citenamefont {Xu},
  \citenamefont {Liu}, \citenamefont {Datta}, \citenamefont {Knee},
  \citenamefont {Lundeen}, \citenamefont {Lu},\ and\ \citenamefont
  {Zhang}}]{Xu20}%
  \BibitemOpen
  \bibfield  {author} {\bibinfo {author} {\bibfnamefont {L.}~\bibnamefont
  {Xu}}, \bibinfo {author} {\bibfnamefont {Z.}~\bibnamefont {Liu}}, \bibinfo
  {author} {\bibfnamefont {A.}~\bibnamefont {Datta}}, \bibinfo {author}
  {\bibfnamefont {G.~C.}\ \bibnamefont {Knee}}, \bibinfo {author}
  {\bibfnamefont {J.~S.}\ \bibnamefont {Lundeen}}, \bibinfo {author}
  {\bibfnamefont {Y.-q.}\ \bibnamefont {Lu}}, \ and\ \bibinfo {author}
  {\bibfnamefont {L.}~\bibnamefont {Zhang}},\ }\href {\doibase
  10.1103/PhysRevLett.125.080501} {\bibfield  {journal} {\bibinfo  {journal}
  {Phys. Rev. Lett.}\ }\textbf {\bibinfo {volume} {125}},\ \bibinfo {pages}
  {080501} (\bibinfo {year} {2020})}\BibitemShut {NoStop}%
\bibitem [{\citenamefont {Arvidsson-Shukur}\ \emph
  {et~al.}(2020{\natexlab{b}})\citenamefont {Arvidsson-Shukur}, \citenamefont
  {Yunger~Halpern}, \citenamefont {Lepage}, \citenamefont {Lasek},
  \citenamefont {Barnes},\ and\ \citenamefont {Lloyd}}]{ArvShukur19-2}%
  \BibitemOpen
  \bibfield  {author} {\bibinfo {author} {\bibfnamefont {D.~R.~M.}\
  \bibnamefont {Arvidsson-Shukur}}, \bibinfo {author} {\bibfnamefont
  {N.}~\bibnamefont {Yunger~Halpern}}, \bibinfo {author} {\bibfnamefont
  {H.~V.}\ \bibnamefont {Lepage}}, \bibinfo {author} {\bibfnamefont {A.~A.}\
  \bibnamefont {Lasek}}, \bibinfo {author} {\bibfnamefont {C.~H.~W.}\
  \bibnamefont {Barnes}}, \ and\ \bibinfo {author} {\bibfnamefont
  {S.}~\bibnamefont {Lloyd}},\ }\href {\doibase 10.1038/s41467-020-17559-w}
  {\bibfield  {journal} {\bibinfo  {journal} {Nature Communications}\ }\textbf
  {\bibinfo {volume} {11}},\ \bibinfo {pages} {3775} (\bibinfo {year}
  {2020}{\natexlab{b}})}\BibitemShut {NoStop}%
\bibitem [{\citenamefont {Lupu-Gladstein~\textit{et al.}}(prep)}]{Lupu20}%
  \BibitemOpen
  \bibfield  {author} {\bibinfo {author} {\bibfnamefont {N.}~\bibnamefont
  {Lupu-Gladstein~\textit{et al.}}},\ }\href@noop {} {\enquote {\bibinfo
  {title} {Experimental demonstration of postselection-enhanced metrology and
  its connection to quantum non-commutation},}\ } (\bibinfo {year} {in
  prep})\BibitemShut {NoStop}%
\bibitem [{\citenamefont {Vella}\ \emph {et~al.}(2019)\citenamefont {Vella},
  \citenamefont {Head}, \citenamefont {Brown},\ and\ \citenamefont
  {Alonso}}]{Vella19}%
  \BibitemOpen
  \bibfield  {author} {\bibinfo {author} {\bibfnamefont {A.}~\bibnamefont
  {Vella}}, \bibinfo {author} {\bibfnamefont {S.~T.}\ \bibnamefont {Head}},
  \bibinfo {author} {\bibfnamefont {T.~G.}\ \bibnamefont {Brown}}, \ and\
  \bibinfo {author} {\bibfnamefont {M.~A.}\ \bibnamefont {Alonso}},\ }\href
  {\doibase 10.1103/PhysRevLett.122.123603} {\bibfield  {journal} {\bibinfo
  {journal} {Phys. Rev. Lett.}\ }\textbf {\bibinfo {volume} {122}},\ \bibinfo
  {pages} {123603} (\bibinfo {year} {2019})}\BibitemShut {NoStop}%
\bibitem [{\citenamefont {Xia}\ \emph {et~al.}(2020)\citenamefont {Xia},
  \citenamefont {Huang}, \citenamefont {Fang}, \citenamefont {Li},\ and\
  \citenamefont {Zeng}}]{Xia20}%
  \BibitemOpen
  \bibfield  {author} {\bibinfo {author} {\bibfnamefont {B.}~\bibnamefont
  {Xia}}, \bibinfo {author} {\bibfnamefont {J.}~\bibnamefont {Huang}}, \bibinfo
  {author} {\bibfnamefont {C.}~\bibnamefont {Fang}}, \bibinfo {author}
  {\bibfnamefont {H.}~\bibnamefont {Li}}, \ and\ \bibinfo {author}
  {\bibfnamefont {G.}~\bibnamefont {Zeng}},\ }\href {\doibase
  10.1103/PhysRevApplied.13.034023} {\bibfield  {journal} {\bibinfo  {journal}
  {Phys. Rev. Applied}\ }\textbf {\bibinfo {volume} {13}},\ \bibinfo {pages}
  {034023} (\bibinfo {year} {2020})}\BibitemShut {NoStop}%
\bibitem [{\citenamefont {Ho}\ and\ \citenamefont {Kondo}(2021)}]{Ho21}%
  \BibitemOpen
  \bibfield  {author} {\bibinfo {author} {\bibfnamefont {L.~B.}\ \bibnamefont
  {Ho}}\ and\ \bibinfo {author} {\bibfnamefont {Y.}~\bibnamefont {Kondo}},\
  }\href {\doibase 10.1063/5.0024555} {\bibfield  {journal} {\bibinfo
  {journal} {Journal of Mathematical Physics}\ }\textbf {\bibinfo {volume}
  {62}},\ \bibinfo {pages} {012102} (\bibinfo {year} {2021})},\ \Eprint
  {http://arxiv.org/abs/https://doi.org/10.1063/5.0024555}
  {https://doi.org/10.1063/5.0024555} \BibitemShut {NoStop}%
\bibitem [{\citenamefont {Kirkwood}(1933)}]{Kirkwood33}%
  \BibitemOpen
  \bibfield  {author} {\bibinfo {author} {\bibfnamefont {J.~G.}\ \bibnamefont
  {Kirkwood}},\ }\href {\doibase 10.1103/PhysRev.44.31} {\bibfield  {journal}
  {\bibinfo  {journal} {Phys. Rev.}\ }\textbf {\bibinfo {volume} {44}},\
  \bibinfo {pages} {31} (\bibinfo {year} {1933})}\BibitemShut {NoStop}%
\bibitem [{\citenamefont {Dirac}(1945)}]{Dirac45}%
  \BibitemOpen
  \bibfield  {author} {\bibinfo {author} {\bibfnamefont {P.~A.~M.}\
  \bibnamefont {Dirac}},\ }\href {\doibase 10.1103/RevModPhys.17.195}
  {\bibfield  {journal} {\bibinfo  {journal} {Rev. Mod. Phys.}\ }\textbf
  {\bibinfo {volume} {17}},\ \bibinfo {pages} {195} (\bibinfo {year}
  {1945})}\BibitemShut {NoStop}%
\bibitem [{\citenamefont {Yunger~Halpern}\ \emph {et~al.}(2018)\citenamefont
  {Yunger~Halpern}, \citenamefont {Swingle},\ and\ \citenamefont
  {Dressel}}]{Yunger18}%
  \BibitemOpen
  \bibfield  {author} {\bibinfo {author} {\bibfnamefont {N.}~\bibnamefont
  {Yunger~Halpern}}, \bibinfo {author} {\bibfnamefont {B.}~\bibnamefont
  {Swingle}}, \ and\ \bibinfo {author} {\bibfnamefont {J.}~\bibnamefont
  {Dressel}},\ }\href {\doibase 10.1103/PhysRevA.97.042105} {\bibfield
  {journal} {\bibinfo  {journal} {Phys. Rev. A}\ }\textbf {\bibinfo {volume}
  {97}},\ \bibinfo {pages} {042105} (\bibinfo {year} {2018})}\BibitemShut
  {NoStop}%
\bibitem [{\citenamefont {Cover}\ and\ \citenamefont
  {Thomas}(2006)}]{bCover06}%
  \BibitemOpen
  \bibfield  {author} {\bibinfo {author} {\bibfnamefont {T.~M.}\ \bibnamefont
  {Cover}}\ and\ \bibinfo {author} {\bibfnamefont {J.~A.}\ \bibnamefont
  {Thomas}},\ }\href@noop {} {\emph {\bibinfo {title} {Elements of Information
  Theory}}},\ \bibinfo {edition} {2nd}\ ed.\ (\bibinfo  {publisher} {John Wiley
  and Sons Inc.},\ \bibinfo {address} {Hoboken, New Jersey, USA},\ \bibinfo
  {year} {2006})\BibitemShut {NoStop}%
\bibitem [{\citenamefont {Lehmann}\ and\ \citenamefont
  {Casella}(2006)}]{Lehmann06}%
  \BibitemOpen
  \bibfield  {author} {\bibinfo {author} {\bibfnamefont {E.~L.}\ \bibnamefont
  {Lehmann}}\ and\ \bibinfo {author} {\bibfnamefont {G.}~\bibnamefont
  {Casella}},\ }\href
  {https://citations.springernature.com/book?doi=10.1007/b98854} {\emph
  {\bibinfo {title} {Theory of point estimation}}}\ (\bibinfo  {publisher}
  {Springer Science \& Business Media},\ \bibinfo {year} {2006})\BibitemShut
  {NoStop}%
\bibitem [{\citenamefont {Helstrom}(1967)}]{Helstrom67}%
  \BibitemOpen
  \bibfield  {author} {\bibinfo {author} {\bibfnamefont {C.}~\bibnamefont
  {Helstrom}},\ }\href {\doibase https://doi.org/10.1016/0375-9601(67)90366-0}
  {\bibfield  {journal} {\bibinfo  {journal} {Physics Letters A}\ }\textbf
  {\bibinfo {volume} {25}},\ \bibinfo {pages} {101} (\bibinfo {year}
  {1967})}\BibitemShut {NoStop}%
\bibitem [{\citenamefont {Abbas}\ \emph {et~al.}(2020)\citenamefont {Abbas},
  \citenamefont {Sutter}, \citenamefont {Zoufal}, \citenamefont {Lucchi},
  \citenamefont {Figalli},\ and\ \citenamefont {Woerner}}]{Abbas20}%
  \BibitemOpen
  \bibfield  {author} {\bibinfo {author} {\bibfnamefont {A.}~\bibnamefont
  {Abbas}}, \bibinfo {author} {\bibfnamefont {D.}~\bibnamefont {Sutter}},
  \bibinfo {author} {\bibfnamefont {C.}~\bibnamefont {Zoufal}}, \bibinfo
  {author} {\bibfnamefont {A.}~\bibnamefont {Lucchi}}, \bibinfo {author}
  {\bibfnamefont {A.}~\bibnamefont {Figalli}}, \ and\ \bibinfo {author}
  {\bibfnamefont {S.}~\bibnamefont {Woerner}},\ }\href
  {https://arxiv.org/abs/2011.00027} {\bibfield  {journal} {\bibinfo  {journal}
  {arXiv preprint arXiv:2011.00027}\ } (\bibinfo {year} {2020})}\BibitemShut
  {NoStop}%
\bibitem [{\citenamefont {Haug}\ \emph {et~al.}(2021)\citenamefont {Haug},
  \citenamefont {Bharti},\ and\ \citenamefont {Kim}}]{Haug21}%
  \BibitemOpen
  \bibfield  {author} {\bibinfo {author} {\bibfnamefont {T.}~\bibnamefont
  {Haug}}, \bibinfo {author} {\bibfnamefont {K.}~\bibnamefont {Bharti}}, \ and\
  \bibinfo {author} {\bibfnamefont {M.}~\bibnamefont {Kim}},\ }\href
  {https://arxiv.org/abs/2102.01659} {\bibfield  {journal} {\bibinfo  {journal}
  {arXiv preprint arXiv:2102.01659}\ } (\bibinfo {year} {2021})}\BibitemShut
  {NoStop}%
\bibitem [{\citenamefont {Meyer}(2021)}]{Meyer21}%
  \BibitemOpen
  \bibfield  {author} {\bibinfo {author} {\bibfnamefont {J.~J.}\ \bibnamefont
  {Meyer}},\ }\href {https://arxiv.org/abs/2103.15191} {\bibfield  {journal}
  {\bibinfo  {journal} {arXiv preprint arXiv:2103.15191}\ } (\bibinfo {year}
  {2021})}\BibitemShut {NoStop}%
\bibitem [{\citenamefont {Albarelli}\ \emph {et~al.}(2019)\citenamefont
  {Albarelli}, \citenamefont {Friel},\ and\ \citenamefont
  {Datta}}]{Albarelli19}%
  \BibitemOpen
  \bibfield  {author} {\bibinfo {author} {\bibfnamefont {F.}~\bibnamefont
  {Albarelli}}, \bibinfo {author} {\bibfnamefont {J.~F.}\ \bibnamefont
  {Friel}}, \ and\ \bibinfo {author} {\bibfnamefont {A.}~\bibnamefont
  {Datta}},\ }\href {\doibase 10.1103/PhysRevLett.123.200503} {\bibfield
  {journal} {\bibinfo  {journal} {Phys. Rev. Lett.}\ }\textbf {\bibinfo
  {volume} {123}},\ \bibinfo {pages} {200503} (\bibinfo {year}
  {2019})}\BibitemShut {NoStop}%
\bibitem [{\citenamefont {Carollo}\ \emph {et~al.}(2019)\citenamefont
  {Carollo}, \citenamefont {Spagnolo}, \citenamefont {Dubkov},\ and\
  \citenamefont {Valenti}}]{Carollo19}%
  \BibitemOpen
  \bibfield  {author} {\bibinfo {author} {\bibfnamefont {A.}~\bibnamefont
  {Carollo}}, \bibinfo {author} {\bibfnamefont {B.}~\bibnamefont {Spagnolo}},
  \bibinfo {author} {\bibfnamefont {A.~A.}\ \bibnamefont {Dubkov}}, \ and\
  \bibinfo {author} {\bibfnamefont {D.}~\bibnamefont {Valenti}},\ }\href
  {https://iopscience.iop.org/article/10.1088/1742-5468/ab3ccb} {\bibfield
  {journal} {\bibinfo  {journal} {Journal of Statistical Mechanics: Theory and
  Experiment}\ }\textbf {\bibinfo {volume} {2019}},\ \bibinfo {pages} {094010}
  (\bibinfo {year} {2019})}\BibitemShut {NoStop}%
\bibitem [{\citenamefont {Holevo}(1977)}]{Holevo77}%
  \BibitemOpen
  \bibfield  {author} {\bibinfo {author} {\bibfnamefont {A.}~\bibnamefont
  {Holevo}},\ }\href {\doibase https://doi.org/10.1016/0034-4877(77)90009-X}
  {\bibfield  {journal} {\bibinfo  {journal} {Reports on Mathematical Physics}\
  }\textbf {\bibinfo {volume} {12}},\ \bibinfo {pages} {251} (\bibinfo {year}
  {1977})}\BibitemShut {NoStop}%
\bibitem [{\citenamefont {Matsumoto}(2002)}]{Matsumoto02}%
  \BibitemOpen
  \bibfield  {author} {\bibinfo {author} {\bibfnamefont {K.}~\bibnamefont
  {Matsumoto}},\ }\href
  {https://iopscience.iop.org/article/10.1088/0305-4470/35/13/307} {\bibfield
  {journal} {\bibinfo  {journal} {Journal of Physics A: Mathematical and
  General}\ }\textbf {\bibinfo {volume} {35}},\ \bibinfo {pages} {3111}
  (\bibinfo {year} {2002})}\BibitemShut {NoStop}%
\bibitem [{\citenamefont {Stone}(1932)}]{Stone32}%
  \BibitemOpen
  \bibfield  {author} {\bibinfo {author} {\bibfnamefont {M.~H.}\ \bibnamefont
  {Stone}},\ }\href@noop {} {\bibfield  {journal} {\bibinfo  {journal} {Ann.
  Math}\ }\textbf {\bibinfo {volume} {33}},\ \bibinfo {pages} {643} (\bibinfo
  {year} {1932})}\BibitemShut {NoStop}%
\bibitem [{\citenamefont {Wigner}(1932)}]{Wigner32}%
  \BibitemOpen
  \bibfield  {author} {\bibinfo {author} {\bibfnamefont {E.}~\bibnamefont
  {Wigner}},\ }\href {\doibase 10.1103/PhysRev.40.749} {\bibfield  {journal}
  {\bibinfo  {journal} {Phys. Rev.}\ }\textbf {\bibinfo {volume} {40}},\
  \bibinfo {pages} {749} (\bibinfo {year} {1932})}\BibitemShut {NoStop}%
\bibitem [{\citenamefont {Steinberg}(1995)}]{Steinberg95}%
  \BibitemOpen
  \bibfield  {author} {\bibinfo {author} {\bibfnamefont {A.~M.}\ \bibnamefont
  {Steinberg}},\ }\href {\doibase 10.1103/PhysRevA.52.32} {\bibfield  {journal}
  {\bibinfo  {journal} {Phys. Rev. A}\ }\textbf {\bibinfo {volume} {52}},\
  \bibinfo {pages} {32} (\bibinfo {year} {1995})}\BibitemShut {NoStop}%
\bibitem [{\citenamefont {Dressel}(2015)}]{Dressel15}%
  \BibitemOpen
  \bibfield  {author} {\bibinfo {author} {\bibfnamefont {J.}~\bibnamefont
  {Dressel}},\ }\href {\doibase 10.1103/PhysRevA.91.032116} {\bibfield
  {journal} {\bibinfo  {journal} {Phys. Rev. A}\ }\textbf {\bibinfo {volume}
  {91}},\ \bibinfo {pages} {032116} (\bibinfo {year} {2015})}\BibitemShut
  {NoStop}%
\bibitem [{\citenamefont {Hofmann}(2011)}]{Hofmann11}%
  \BibitemOpen
  \bibfield  {author} {\bibinfo {author} {\bibfnamefont {H.~F.}\ \bibnamefont
  {Hofmann}},\ }\href {\doibase 10.1088/1367-2630/13/10/103009} {\bibfield
  {journal} {\bibinfo  {journal} {New Journal of Physics}\ }\textbf {\bibinfo
  {volume} {13}},\ \bibinfo {pages} {103009} (\bibinfo {year}
  {2011})}\BibitemShut {NoStop}%
\bibitem [{\citenamefont {Dressel}\ and\ \citenamefont
  {Jordan}(2012)}]{Dressel12}%
  \BibitemOpen
  \bibfield  {author} {\bibinfo {author} {\bibfnamefont {J.}~\bibnamefont
  {Dressel}}\ and\ \bibinfo {author} {\bibfnamefont {A.~N.}\ \bibnamefont
  {Jordan}},\ }\href {\doibase 10.1103/PhysRevA.85.012107} {\bibfield
  {journal} {\bibinfo  {journal} {Phys. Rev. A}\ }\textbf {\bibinfo {volume}
  {85}},\ \bibinfo {pages} {012107} (\bibinfo {year} {2012})}\BibitemShut
  {NoStop}%
\bibitem [{\citenamefont {Monroe}\ \emph {et~al.}(2020)\citenamefont {Monroe},
  \citenamefont {Halpern}, \citenamefont {Lee},\ and\ \citenamefont
  {Murch}}]{Monroe20}%
  \BibitemOpen
  \bibfield  {author} {\bibinfo {author} {\bibfnamefont {J.~T.}\ \bibnamefont
  {Monroe}}, \bibinfo {author} {\bibfnamefont {N.~Y.}\ \bibnamefont {Halpern}},
  \bibinfo {author} {\bibfnamefont {T.}~\bibnamefont {Lee}}, \ and\ \bibinfo
  {author} {\bibfnamefont {K.~W.}\ \bibnamefont {Murch}},\ }\href
  {https://arxiv.org/abs/2008.09131} {\bibfield  {journal} {\bibinfo  {journal}
  {arXiv preprint arXiv:2008.09131}\ } (\bibinfo {year} {2020})}\BibitemShut
  {NoStop}%
\bibitem [{\citenamefont {Johansen}(2007)}]{Johansen07}%
  \BibitemOpen
  \bibfield  {author} {\bibinfo {author} {\bibfnamefont {L.~M.}\ \bibnamefont
  {Johansen}},\ }\href {\doibase 10.1103/PhysRevA.76.012119} {\bibfield
  {journal} {\bibinfo  {journal} {Phys. Rev. A}\ }\textbf {\bibinfo {volume}
  {76}},\ \bibinfo {pages} {012119} (\bibinfo {year} {2007})}\BibitemShut
  {NoStop}%
\bibitem [{\citenamefont {Lundeen}\ \emph {et~al.}(2011)\citenamefont
  {Lundeen}, \citenamefont {Sutherland}, \citenamefont {Patel}, \citenamefont
  {Stewart},\ and\ \citenamefont {Bamber}}]{Lundeen11}%
  \BibitemOpen
  \bibfield  {author} {\bibinfo {author} {\bibfnamefont {J.~S.}\ \bibnamefont
  {Lundeen}}, \bibinfo {author} {\bibfnamefont {B.}~\bibnamefont {Sutherland}},
  \bibinfo {author} {\bibfnamefont {A.}~\bibnamefont {Patel}}, \bibinfo
  {author} {\bibfnamefont {C.}~\bibnamefont {Stewart}}, \ and\ \bibinfo
  {author} {\bibfnamefont {C.}~\bibnamefont {Bamber}},\ }\href
  {https://doi.org/10.1038/nature10120} {\bibfield  {journal} {\bibinfo
  {journal} {Nature}\ }\textbf {\bibinfo {volume} {474}},\ \bibinfo {pages}
  {188} (\bibinfo {year} {2011})}\BibitemShut {NoStop}%
\bibitem [{\citenamefont {Lundeen}\ and\ \citenamefont
  {Bamber}(2012)}]{Lundeen12}%
  \BibitemOpen
  \bibfield  {author} {\bibinfo {author} {\bibfnamefont {J.~S.}\ \bibnamefont
  {Lundeen}}\ and\ \bibinfo {author} {\bibfnamefont {C.}~\bibnamefont
  {Bamber}},\ }\href {\doibase 10.1103/PhysRevLett.108.070402} {\bibfield
  {journal} {\bibinfo  {journal} {Phys. Rev. Lett.}\ }\textbf {\bibinfo
  {volume} {108}},\ \bibinfo {pages} {070402} (\bibinfo {year}
  {2012})}\BibitemShut {NoStop}%
\bibitem [{\citenamefont {Bamber}\ and\ \citenamefont
  {Lundeen}(2014)}]{Bamber14}%
  \BibitemOpen
  \bibfield  {author} {\bibinfo {author} {\bibfnamefont {C.}~\bibnamefont
  {Bamber}}\ and\ \bibinfo {author} {\bibfnamefont {J.~S.}\ \bibnamefont
  {Lundeen}},\ }\href {\doibase 10.1103/PhysRevLett.112.070405} {\bibfield
  {journal} {\bibinfo  {journal} {Phys. Rev. Lett.}\ }\textbf {\bibinfo
  {volume} {112}},\ \bibinfo {pages} {070405} (\bibinfo {year}
  {2014})}\BibitemShut {NoStop}%
\bibitem [{\citenamefont {Thekkadath}\ \emph {et~al.}(2016)\citenamefont
  {Thekkadath}, \citenamefont {Giner}, \citenamefont {Chalich}, \citenamefont
  {Horton}, \citenamefont {Banker},\ and\ \citenamefont
  {Lundeen}}]{Thekkadath16}%
  \BibitemOpen
  \bibfield  {author} {\bibinfo {author} {\bibfnamefont {G.~S.}\ \bibnamefont
  {Thekkadath}}, \bibinfo {author} {\bibfnamefont {L.}~\bibnamefont {Giner}},
  \bibinfo {author} {\bibfnamefont {Y.}~\bibnamefont {Chalich}}, \bibinfo
  {author} {\bibfnamefont {M.~J.}\ \bibnamefont {Horton}}, \bibinfo {author}
  {\bibfnamefont {J.}~\bibnamefont {Banker}}, \ and\ \bibinfo {author}
  {\bibfnamefont {J.~S.}\ \bibnamefont {Lundeen}},\ }\href {\doibase
  10.1103/PhysRevLett.117.120401} {\bibfield  {journal} {\bibinfo  {journal}
  {Phys. Rev. Lett.}\ }\textbf {\bibinfo {volume} {117}},\ \bibinfo {pages}
  {120401} (\bibinfo {year} {2016})}\BibitemShut {NoStop}%
\bibitem [{\citenamefont {Swingle}\ \emph {et~al.}(2016)\citenamefont
  {Swingle}, \citenamefont {Bentsen}, \citenamefont {Schleier-Smith},\ and\
  \citenamefont {Hayden}}]{Swingle16}%
  \BibitemOpen
  \bibfield  {author} {\bibinfo {author} {\bibfnamefont {B.}~\bibnamefont
  {Swingle}}, \bibinfo {author} {\bibfnamefont {G.}~\bibnamefont {Bentsen}},
  \bibinfo {author} {\bibfnamefont {M.}~\bibnamefont {Schleier-Smith}}, \ and\
  \bibinfo {author} {\bibfnamefont {P.}~\bibnamefont {Hayden}},\ }\href
  {\doibase 10.1103/PhysRevA.94.040302} {\bibfield  {journal} {\bibinfo
  {journal} {Phys. Rev. A}\ }\textbf {\bibinfo {volume} {94}},\ \bibinfo
  {pages} {040302} (\bibinfo {year} {2016})}\BibitemShut {NoStop}%
\bibitem [{\citenamefont {Halpern}\ \emph {et~al.}(2019)\citenamefont
  {Halpern}, \citenamefont {Bartolotta},\ and\ \citenamefont
  {Pollack}}]{Yunger18-2}%
  \BibitemOpen
  \bibfield  {author} {\bibinfo {author} {\bibfnamefont {N.~Y.}\ \bibnamefont
  {Halpern}}, \bibinfo {author} {\bibfnamefont {A.}~\bibnamefont {Bartolotta}},
  \ and\ \bibinfo {author} {\bibfnamefont {J.}~\bibnamefont {Pollack}},\ }\href
  {\doibase 10.1038/s42005-019-0179-8} {\bibfield  {journal} {\bibinfo
  {journal} {Communications Physics}\ }\textbf {\bibinfo {volume} {2}},\
  \bibinfo {pages} {1} (\bibinfo {year} {2019})}\BibitemShut {NoStop}%
\bibitem [{\citenamefont {Gonz\'alez~Alonso}\ \emph {et~al.}(2019)\citenamefont
  {Gonz\'alez~Alonso}, \citenamefont {Yunger~Halpern},\ and\ \citenamefont
  {Dressel}}]{Yunger19}%
  \BibitemOpen
  \bibfield  {author} {\bibinfo {author} {\bibfnamefont {J.~R.}\ \bibnamefont
  {Gonz\'alez~Alonso}}, \bibinfo {author} {\bibfnamefont {N.}~\bibnamefont
  {Yunger~Halpern}}, \ and\ \bibinfo {author} {\bibfnamefont {J.}~\bibnamefont
  {Dressel}},\ }\href {\doibase 10.1103/PhysRevLett.122.040404} {\bibfield
  {journal} {\bibinfo  {journal} {Phys. Rev. Lett.}\ }\textbf {\bibinfo
  {volume} {122}},\ \bibinfo {pages} {040404} (\bibinfo {year}
  {2019})}\BibitemShut {NoStop}%
\bibitem [{\citenamefont {Landsman}\ \emph {et~al.}(2019)\citenamefont
  {Landsman}, \citenamefont {Figgatt}, \citenamefont {Schuster}, \citenamefont
  {Linke}, \citenamefont {Yoshida}, \citenamefont {Yao},\ and\ \citenamefont
  {Monroe}}]{Landsman19}%
  \BibitemOpen
  \bibfield  {author} {\bibinfo {author} {\bibfnamefont {K.~A.}\ \bibnamefont
  {Landsman}}, \bibinfo {author} {\bibfnamefont {C.}~\bibnamefont {Figgatt}},
  \bibinfo {author} {\bibfnamefont {T.}~\bibnamefont {Schuster}}, \bibinfo
  {author} {\bibfnamefont {N.~M.}\ \bibnamefont {Linke}}, \bibinfo {author}
  {\bibfnamefont {B.}~\bibnamefont {Yoshida}}, \bibinfo {author} {\bibfnamefont
  {N.~Y.}\ \bibnamefont {Yao}}, \ and\ \bibinfo {author} {\bibfnamefont
  {C.}~\bibnamefont {Monroe}},\ }\href {\doibase
  https://doi.org/10.1038/s41586-019-0952-6} {\bibfield  {journal} {\bibinfo
  {journal} {Nature}\ }\textbf {\bibinfo {volume} {567}},\ \bibinfo {pages}
  {61} (\bibinfo {year} {2019})}\BibitemShut {NoStop}%
\bibitem [{\citenamefont {Mohseninia}\ \emph {et~al.}(2019)\citenamefont
  {Mohseninia}, \citenamefont {Alonso},\ and\ \citenamefont
  {Dressel}}]{Razieh19}%
  \BibitemOpen
  \bibfield  {author} {\bibinfo {author} {\bibfnamefont {R.}~\bibnamefont
  {Mohseninia}}, \bibinfo {author} {\bibfnamefont {J.~R.~G.}\ \bibnamefont
  {Alonso}}, \ and\ \bibinfo {author} {\bibfnamefont {J.}~\bibnamefont
  {Dressel}},\ }\href {\doibase 10.1103/PhysRevA.100.062336} {\bibfield
  {journal} {\bibinfo  {journal} {Phys. Rev. A}\ }\textbf {\bibinfo {volume}
  {100}},\ \bibinfo {pages} {062336} (\bibinfo {year} {2019})}\BibitemShut
  {NoStop}%
\bibitem [{\citenamefont {Levy}\ and\ \citenamefont
  {Lostaglio}(2019)}]{Levy19}%
  \BibitemOpen
  \bibfield  {author} {\bibinfo {author} {\bibfnamefont {A.}~\bibnamefont
  {Levy}}\ and\ \bibinfo {author} {\bibfnamefont {M.}~\bibnamefont
  {Lostaglio}},\ }\href {https://arxiv.org/abs/1909.11116} {\bibfield
  {journal} {\bibinfo  {journal} {arXiv preprint arXiv:1909.11116}\ } (\bibinfo
  {year} {2019})}\BibitemShut {NoStop}%
\bibitem [{\citenamefont {Lostaglio}(2020)}]{Lostaglio20}%
  \BibitemOpen
  \bibfield  {author} {\bibinfo {author} {\bibfnamefont {M.}~\bibnamefont
  {Lostaglio}},\ }\href {https://arxiv.org/abs/2004.01213} {\bibfield
  {journal} {\bibinfo  {journal} {arXiv preprint arXiv:2004.01213}\ } (\bibinfo
  {year} {2020})}\BibitemShut {NoStop}%
\bibitem [{\citenamefont {Griffiths}(1984)}]{Griffiths84}%
  \BibitemOpen
  \bibfield  {author} {\bibinfo {author} {\bibfnamefont {R.~B.}\ \bibnamefont
  {Griffiths}},\ }\href {\doibase https://doi.org/10.1007/BF01015734}
  {\bibfield  {journal} {\bibinfo  {journal} {Journal of Statistical Physics}\
  }\textbf {\bibinfo {volume} {36}},\ \bibinfo {pages} {219} (\bibinfo {year}
  {1984})}\BibitemShut {NoStop}%
\bibitem [{\citenamefont {Goldstein}\ and\ \citenamefont
  {Page}(1995)}]{Goldstein95}%
  \BibitemOpen
  \bibfield  {author} {\bibinfo {author} {\bibfnamefont {S.}~\bibnamefont
  {Goldstein}}\ and\ \bibinfo {author} {\bibfnamefont {D.~N.}\ \bibnamefont
  {Page}},\ }\href {\doibase 10.1103/PhysRevLett.74.3715} {\bibfield  {journal}
  {\bibinfo  {journal} {Phys. Rev. Lett.}\ }\textbf {\bibinfo {volume} {74}},\
  \bibinfo {pages} {3715} (\bibinfo {year} {1995})}\BibitemShut {NoStop}%
\bibitem [{\citenamefont {Hartle}(2004)}]{Hartle04}%
  \BibitemOpen
  \bibfield  {author} {\bibinfo {author} {\bibfnamefont {J.~B.}\ \bibnamefont
  {Hartle}},\ }\href {\doibase 10.1103/PhysRevA.70.022104} {\bibfield
  {journal} {\bibinfo  {journal} {Phys. Rev. A}\ }\textbf {\bibinfo {volume}
  {70}},\ \bibinfo {pages} {022104} (\bibinfo {year} {2004})}\BibitemShut
  {NoStop}%
\bibitem [{\citenamefont {Hofmann}(2012)}]{Hofmann12-2}%
  \BibitemOpen
  \bibfield  {author} {\bibinfo {author} {\bibfnamefont {H.~F.}\ \bibnamefont
  {Hofmann}},\ }\href {\doibase 10.1088/1367-2630/14/4/043031} {\bibfield
  {journal} {\bibinfo  {journal} {New Journal of Physics}\ }\textbf {\bibinfo
  {volume} {14}},\ \bibinfo {pages} {043031} (\bibinfo {year}
  {2012})}\BibitemShut {NoStop}%
\bibitem [{\citenamefont {Hofmann}(2014)}]{Hofmann14}%
  \BibitemOpen
  \bibfield  {author} {\bibinfo {author} {\bibfnamefont {H.~F.}\ \bibnamefont
  {Hofmann}},\ }\href {\doibase 10.1103/PhysRevA.89.042115} {\bibfield
  {journal} {\bibinfo  {journal} {Phys. Rev. A}\ }\textbf {\bibinfo {volume}
  {89}},\ \bibinfo {pages} {042115} (\bibinfo {year} {2014})}\BibitemShut
  {NoStop}%
\bibitem [{\citenamefont {Hofmann}(2015)}]{Hofmann15}%
  \BibitemOpen
  \bibfield  {author} {\bibinfo {author} {\bibfnamefont {H.~F.}\ \bibnamefont
  {Hofmann}},\ }\href {\doibase 10.1103/PhysRevA.91.062123} {\bibfield
  {journal} {\bibinfo  {journal} {Phys. Rev. A}\ }\textbf {\bibinfo {volume}
  {91}},\ \bibinfo {pages} {062123} (\bibinfo {year} {2015})}\BibitemShut
  {NoStop}%
\bibitem [{\citenamefont {Hofmann}(2016)}]{Hofmann16}%
  \BibitemOpen
  \bibfield  {author} {\bibinfo {author} {\bibfnamefont {H.~F.}\ \bibnamefont
  {Hofmann}},\ }\href {\doibase https://doi.org/10.1140/epjd/e2016-70086-8}
  {\bibfield  {journal} {\bibinfo  {journal} {The European Physical Journal D}\
  }\textbf {\bibinfo {volume} {70}},\ \bibinfo {pages} {118} (\bibinfo {year}
  {2016})}\BibitemShut {NoStop}%
\bibitem [{\citenamefont {Halliwell}(2016)}]{Halliwell16}%
  \BibitemOpen
  \bibfield  {author} {\bibinfo {author} {\bibfnamefont {J.~J.}\ \bibnamefont
  {Halliwell}},\ }\href {\doibase 10.1103/PhysRevA.93.022123} {\bibfield
  {journal} {\bibinfo  {journal} {Phys. Rev. A}\ }\textbf {\bibinfo {volume}
  {93}},\ \bibinfo {pages} {022123} (\bibinfo {year} {2016})}\BibitemShut
  {NoStop}%
\bibitem [{\citenamefont {Stacey}(2019)}]{Stacey19}%
  \BibitemOpen
  \bibfield  {author} {\bibinfo {author} {\bibfnamefont {B.~C.}\ \bibnamefont
  {Stacey}},\ }\href {https://arxiv.org/pdf/1907.02432} {\bibfield  {journal}
  {\bibinfo  {journal} {arXiv preprint arXiv:1907.02432}\ } (\bibinfo {year}
  {2019})}\BibitemShut {NoStop}%
\bibitem [{\citenamefont {Tang}\ \emph {et~al.}(2019)\citenamefont {Tang},
  \citenamefont {Shkolnikov}, \citenamefont {Barron}, \citenamefont {Grimsley},
  \citenamefont {Mayhall}, \citenamefont {Barnes},\ and\ \citenamefont
  {Economou}}]{Tang19}%
  \BibitemOpen
  \bibfield  {author} {\bibinfo {author} {\bibfnamefont {H.~L.}\ \bibnamefont
  {Tang}}, \bibinfo {author} {\bibfnamefont {V.}~\bibnamefont {Shkolnikov}},
  \bibinfo {author} {\bibfnamefont {G.~S.}\ \bibnamefont {Barron}}, \bibinfo
  {author} {\bibfnamefont {H.~R.}\ \bibnamefont {Grimsley}}, \bibinfo {author}
  {\bibfnamefont {N.~J.}\ \bibnamefont {Mayhall}}, \bibinfo {author}
  {\bibfnamefont {E.}~\bibnamefont {Barnes}}, \ and\ \bibinfo {author}
  {\bibfnamefont {S.~E.}\ \bibnamefont {Economou}},\ }\href
  {https://arxiv.org/abs/1911.10205} {\bibfield  {journal} {\bibinfo  {journal}
  {arXiv preprint arXiv:1911.10205}\ } (\bibinfo {year} {2019})}\BibitemShut
  {NoStop}%
\bibitem [{\citenamefont {Grimsley}\ \emph {et~al.}(2019)\citenamefont
  {Grimsley}, \citenamefont {Economou}, \citenamefont {Barnes},\ and\
  \citenamefont {Mayhall}}]{Grimsley19}%
  \BibitemOpen
  \bibfield  {author} {\bibinfo {author} {\bibfnamefont {H.~R.}\ \bibnamefont
  {Grimsley}}, \bibinfo {author} {\bibfnamefont {S.~E.}\ \bibnamefont
  {Economou}}, \bibinfo {author} {\bibfnamefont {E.}~\bibnamefont {Barnes}}, \
  and\ \bibinfo {author} {\bibfnamefont {N.~J.}\ \bibnamefont {Mayhall}},\
  }\href {https://www.nature.com/articles/s41467-019-10988-2} {\bibfield
  {journal} {\bibinfo  {journal} {Nature communications}\ }\textbf {\bibinfo
  {volume} {10}},\ \bibinfo {pages} {1} (\bibinfo {year} {2019})}\BibitemShut
  {NoStop}%
\bibitem [{\citenamefont {Yordanov}\ \emph {et~al.}(2020)\citenamefont
  {Yordanov}, \citenamefont {Armaos}, \citenamefont {Barnes},\ and\
  \citenamefont {Arvidsson-Shukur}}]{Yordanov20}%
  \BibitemOpen
  \bibfield  {author} {\bibinfo {author} {\bibfnamefont {Y.~S.}\ \bibnamefont
  {Yordanov}}, \bibinfo {author} {\bibfnamefont {V.}~\bibnamefont {Armaos}},
  \bibinfo {author} {\bibfnamefont {C.~H.}\ \bibnamefont {Barnes}}, \ and\
  \bibinfo {author} {\bibfnamefont {D.~R.}\ \bibnamefont {Arvidsson-Shukur}},\
  }\href {https://arxiv.org/abs/2011.10540} {\bibfield  {journal} {\bibinfo
  {journal} {arXiv preprint arXiv:2011.10540}\ } (\bibinfo {year}
  {2020})}\BibitemShut {NoStop}%
\bibitem [{\citenamefont {McArdle}\ \emph {et~al.}(2020)\citenamefont
  {McArdle}, \citenamefont {Endo}, \citenamefont {Aspuru-Guzik}, \citenamefont
  {Benjamin},\ and\ \citenamefont {Yuan}}]{Benjamin20}%
  \BibitemOpen
  \bibfield  {author} {\bibinfo {author} {\bibfnamefont {S.}~\bibnamefont
  {McArdle}}, \bibinfo {author} {\bibfnamefont {S.}~\bibnamefont {Endo}},
  \bibinfo {author} {\bibfnamefont {A.}~\bibnamefont {Aspuru-Guzik}}, \bibinfo
  {author} {\bibfnamefont {S.~C.}\ \bibnamefont {Benjamin}}, \ and\ \bibinfo
  {author} {\bibfnamefont {X.}~\bibnamefont {Yuan}},\ }\href {\doibase
  10.1103/RevModPhys.92.015003} {\bibfield  {journal} {\bibinfo  {journal}
  {Rev. Mod. Phys.}\ }\textbf {\bibinfo {volume} {92}},\ \bibinfo {pages}
  {015003} (\bibinfo {year} {2020})}\BibitemShut {NoStop}%
\bibitem [{\citenamefont {Lavrijsen}\ \emph {et~al.}(2020)\citenamefont
  {Lavrijsen}, \citenamefont {Tudor}, \citenamefont {M{\"u}ller}, \citenamefont
  {Iancu},\ and\ \citenamefont {de~Jong}}]{Lavrijsen20}%
  \BibitemOpen
  \bibfield  {author} {\bibinfo {author} {\bibfnamefont {W.}~\bibnamefont
  {Lavrijsen}}, \bibinfo {author} {\bibfnamefont {A.}~\bibnamefont {Tudor}},
  \bibinfo {author} {\bibfnamefont {J.}~\bibnamefont {M{\"u}ller}}, \bibinfo
  {author} {\bibfnamefont {C.}~\bibnamefont {Iancu}}, \ and\ \bibinfo {author}
  {\bibfnamefont {W.}~\bibnamefont {de~Jong}},\ }in\ \href
  {https://ieeexplore.ieee.org/document/9259985} {\emph {\bibinfo {booktitle}
  {2020 IEEE International Conference on Quantum Computing and Engineering
  (QCE)}}}\ (\bibinfo {organization} {IEEE},\ \bibinfo {year} {2020})\ pp.\
  \bibinfo {pages} {267--277}\BibitemShut {NoStop}%
\bibitem [{\citenamefont {Bittel}\ and\ \citenamefont
  {Kliesch}(2021)}]{Bittel21}%
  \BibitemOpen
  \bibfield  {author} {\bibinfo {author} {\bibfnamefont {L.}~\bibnamefont
  {Bittel}}\ and\ \bibinfo {author} {\bibfnamefont {M.}~\bibnamefont
  {Kliesch}},\ }\href {https://arxiv.org/abs/2101.07267} {\bibfield  {journal}
  {\bibinfo  {journal} {arXiv preprint arXiv:2101.07267}\ } (\bibinfo {year}
  {2021})}\BibitemShut {NoStop}%
\bibitem [{\citenamefont {Zarchan}\ \emph {et~al.}(2000)\citenamefont
  {Zarchan}, \citenamefont {Musoff}, \citenamefont {of~Aeronautics},\ and\
  \citenamefont {Astronautics}}]{Zarchan00}%
  \BibitemOpen
  \bibfield  {author} {\bibinfo {author} {\bibfnamefont {P.}~\bibnamefont
  {Zarchan}}, \bibinfo {author} {\bibfnamefont {H.}~\bibnamefont {Musoff}},
  \bibinfo {author} {\bibfnamefont {A.~I.}\ \bibnamefont {of~Aeronautics}}, \
  and\ \bibinfo {author} {\bibnamefont {Astronautics}},\ }\href
  {https://books.google.se/books?id=AQxRAAAAMAAJ} {\emph {\bibinfo {title}
  {Fundamentals of Kalman Filtering: A Practical Approach}}},\ Progress in
  astronautics and aeronautics\ (\bibinfo  {publisher} {American Institute of
  Aeronautics and Astronautics, Incorporated},\ \bibinfo {year}
  {2000})\BibitemShut {NoStop}%
\bibitem [{\citenamefont {Carollo}\ \emph {et~al.}(2018)\citenamefont
  {Carollo}, \citenamefont {Spagnolo},\ and\ \citenamefont
  {Valenti}}]{Carollo18}%
  \BibitemOpen
  \bibfield  {author} {\bibinfo {author} {\bibfnamefont {A.}~\bibnamefont
  {Carollo}}, \bibinfo {author} {\bibfnamefont {B.}~\bibnamefont {Spagnolo}}, \
  and\ \bibinfo {author} {\bibfnamefont {D.}~\bibnamefont {Valenti}},\ }\href
  {https://www.nature.com/articles/s41598-018-27362-9} {\bibfield  {journal}
  {\bibinfo  {journal} {Scientific reports}\ }\textbf {\bibinfo {volume} {8}},\
  \bibinfo {pages} {1} (\bibinfo {year} {2018})}\BibitemShut {NoStop}%
\end{thebibliography}%

\end{document}